\documentclass[preprint,11pt]{aastex}
\usepackage{ifthen}
\newboolean{apj}
\setboolean{apj}{true}  

\def\Gyr{{\rm\,Gyr}}
\def\kms{{\rm\,km\,s^{-1}}}
\def\mpc{{\rm\,Mpc}}
\def\yr{{\rm\,yr}}
\def\hmpc{h^{-1}{\rm Mpc}}
\def\pdrv#1#2{{\partial #1 \over \partial #2}}
\def\hinv{{h^{-1}}}
\def\K{{\rm K}}

\begin{document}

\title{The growth of galaxies in cosmological simulations of structure
	formation}

\author{Chigurupati Murali, Neal Katz}
\affil{Astronomy Department, University of Massachusetts, Amherst, MA 01003}
\author{Lars Hernquist}
\affil{Department of Astronomy, Harvard University, Cambridge, MA 02138}
\author{David H. Weinberg}
\affil{Astronomy Department, Ohio State University, Columbus, OH 43210}
\author{Romeel Dav\'e}
\affil{Astrophysical Sciences, Princeton University, Princeton, NJ 08544}

\begin{abstract}
We use hydrodynamic simulations to examine how the baryonic components of 
galaxies are assembled, focusing on the relative importance of mergers and
smooth accretion in the formation of $\sim L_*$ systems.  In our primary
simulation, which models a $(50\hmpc)^3$ comoving volume of a 
$\Lambda$-dominated cold dark matter universe, the space density of objects
at our (64-particle) baryon mass resolution threshold $M_c=5.4\times 10^{10}
M_\odot$ corresponds to that of observed galaxies with $L\sim L_*/4$.  Galaxies
above this threshold gain most of their mass by accretion rather than by 
mergers.  At the redshift of peak mass growth, $z\approx 2$, accretion 
dominates over merging by about 4:1.  The mean accretion rate per galaxy 
declines 
from $\sim 40M_\odot\yr^{-1}$ at $z=2$ to $\sim 10M_\odot\yr^{-1}$ at $z=0$,
while the merging rate peaks later ($z\approx 1$) and declines more slowly,
so by $z=0$ the ratio is about 2:1.  We cannot distinguish truly smooth 
accretion from merging with objects below our mass resolution threshold, but
extrapolating our measured mass spectrum of merging objects, $dP/dM \propto
M^{-\alpha}$ with $\alpha \sim 1$, implies that sub-resolution mergers would 
add relatively little mass.  The global star formation history in these 
simulations tracks the mass accretion rate rather than the merger rate.  At low
redshift, destruction of galaxies by mergers is approximately balanced by the
growth of new systems, so the comoving space density of resolved galaxies stays
nearly constant despite significant mass evolution at the galaxy-by-galaxy 
level.
The predicted merger rate at $z \la 1$ agrees with recent estimates from 
close pairs in the CFRS and CNOC2 redshift surveys.
\end{abstract}



\newcommand{\massfuncfig}{
\begin{figure*}
\plotone{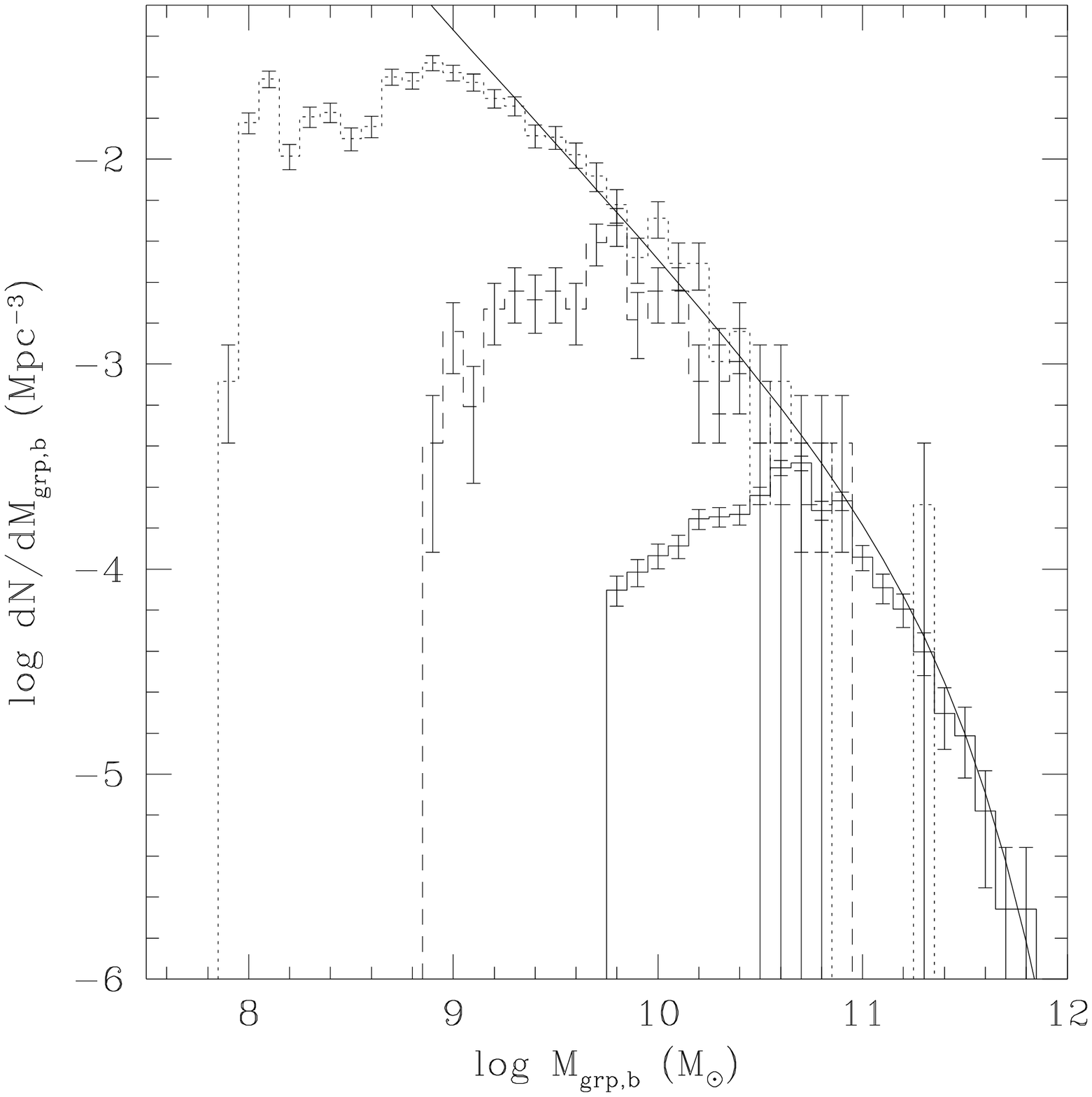}
\caption{
\label{fig.massfunc}
The mass function at $z=3$ in the L50/144 (solid
histogram), L11/64 (dashed histogram), and L11/128 simulations
(dotted histogram).  Error bars reflect Poisson errors. The solid
curve is a Schechter function overlaid for comparison.}
\end{figure*}
}

\newcommand{\densityfig}{
\begin{figure*}
\plotone{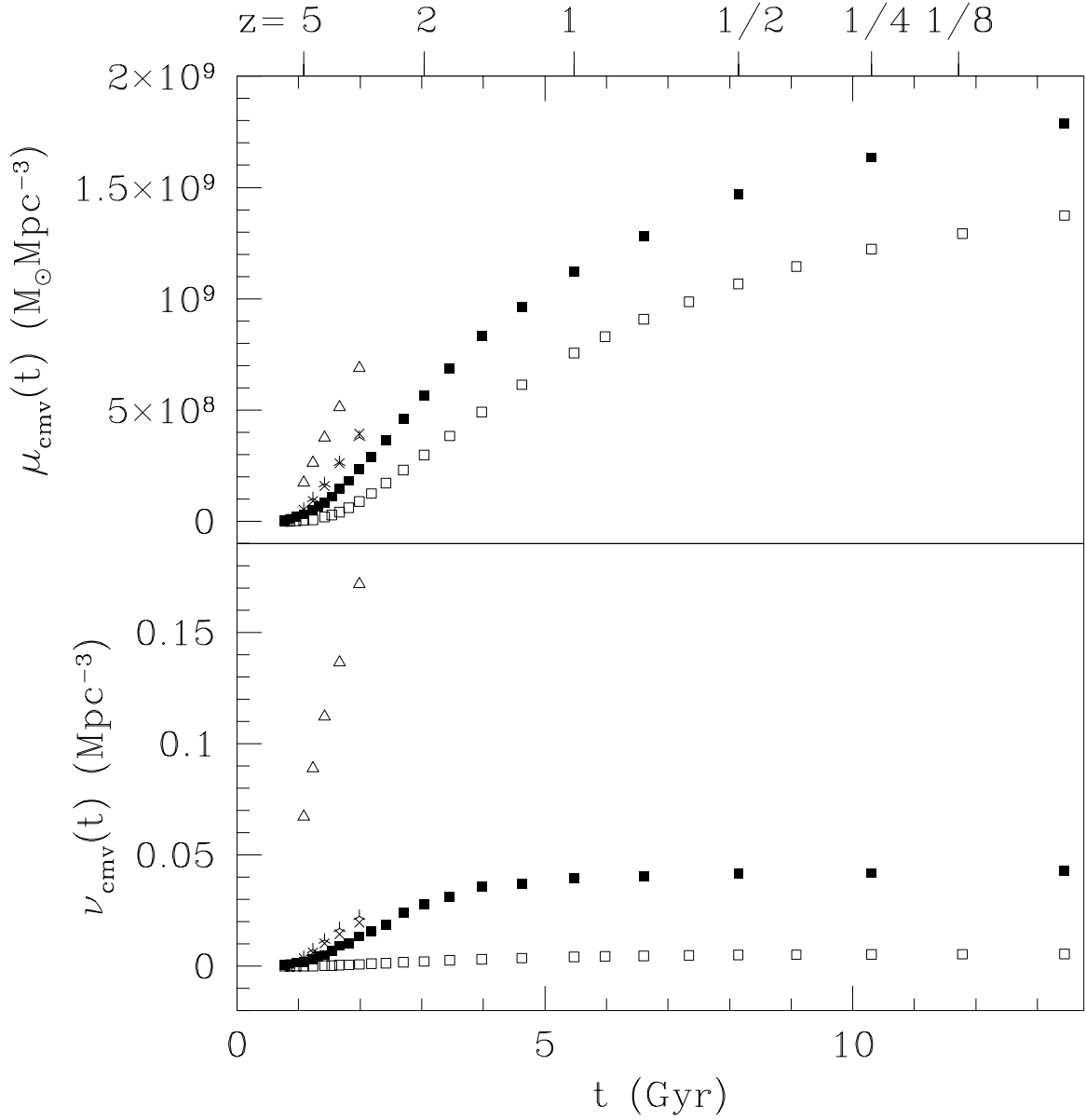}
\caption{
\label{fig.density}
The total comoving mass (top) and number density (bottom) of galaxies
as a function of cosmic time for galaxies with masses above
$M_c = 5.4\times 10^{10} M_\odot$ (L50/144--open squares), 
$M_c = 6.8\times 10^{9} M_\odot$ (L11/64--solid squares, L11/64$^\prime$--four point
stars, L11/128--three point stars), and 
$M_c = 8.5\times 10^{8} M_\odot$ (L11/128--open triangles).}
\end{figure*}
}

\newcommand{\numdenfig}{
\begin{figure*}
\plotone{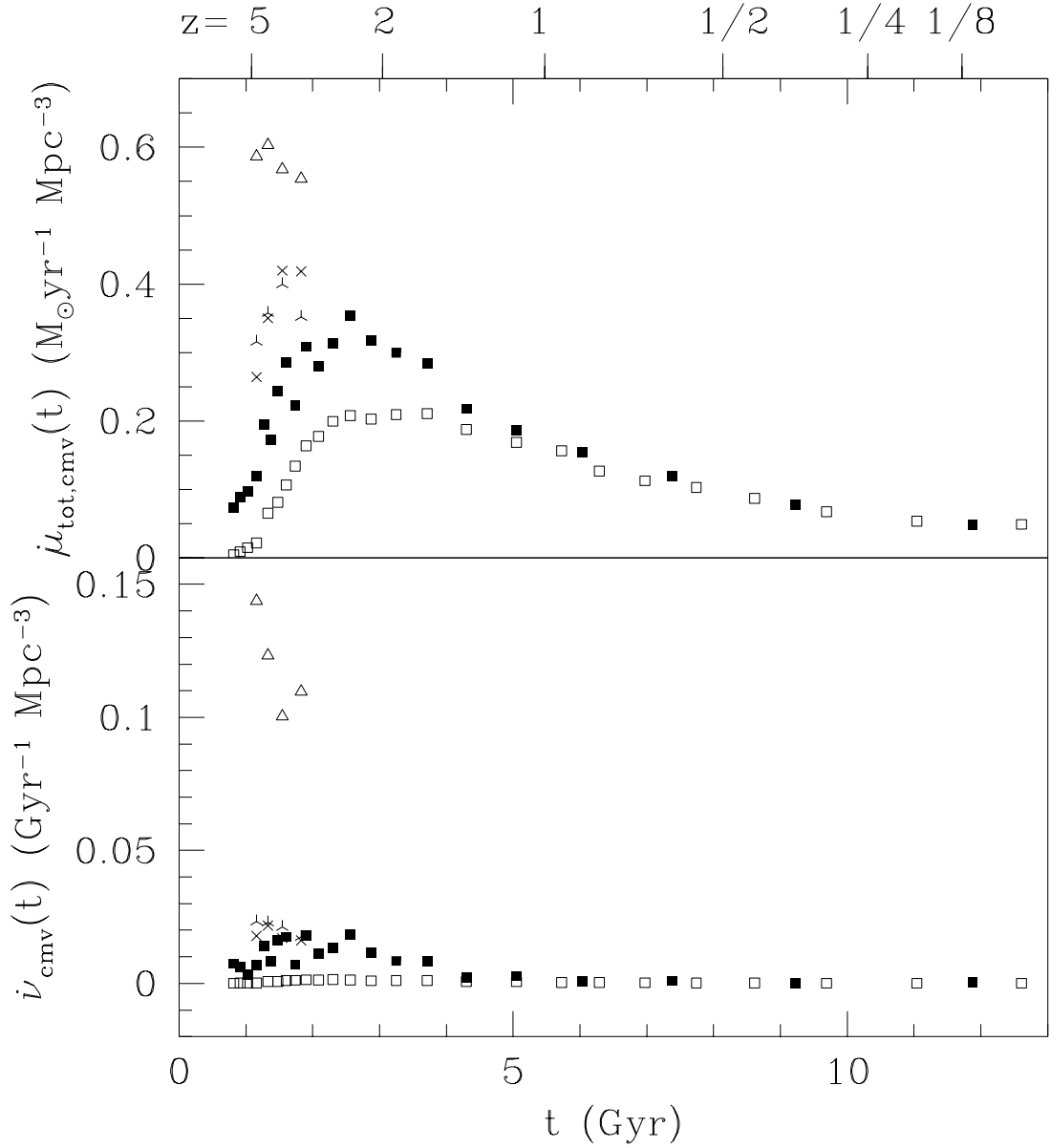}
\caption{
\label{fig.numden}
The rates of change of mass (top) and number density (bottom) above the
masses shown in Figure~\ref{fig.density}.  The number and mass densities
for masses above $M_c = 5.4\times 10^{10} M_\odot$ (L50/144--open squares)
and $M_c = 6.8\times 10^{9} M_\odot$ (L11/64--solid squares) agree 
for $z<1.5$.}
\end{figure*}
}
 
\newcommand{\cratefig}{
\begin{figure*}
\plotone{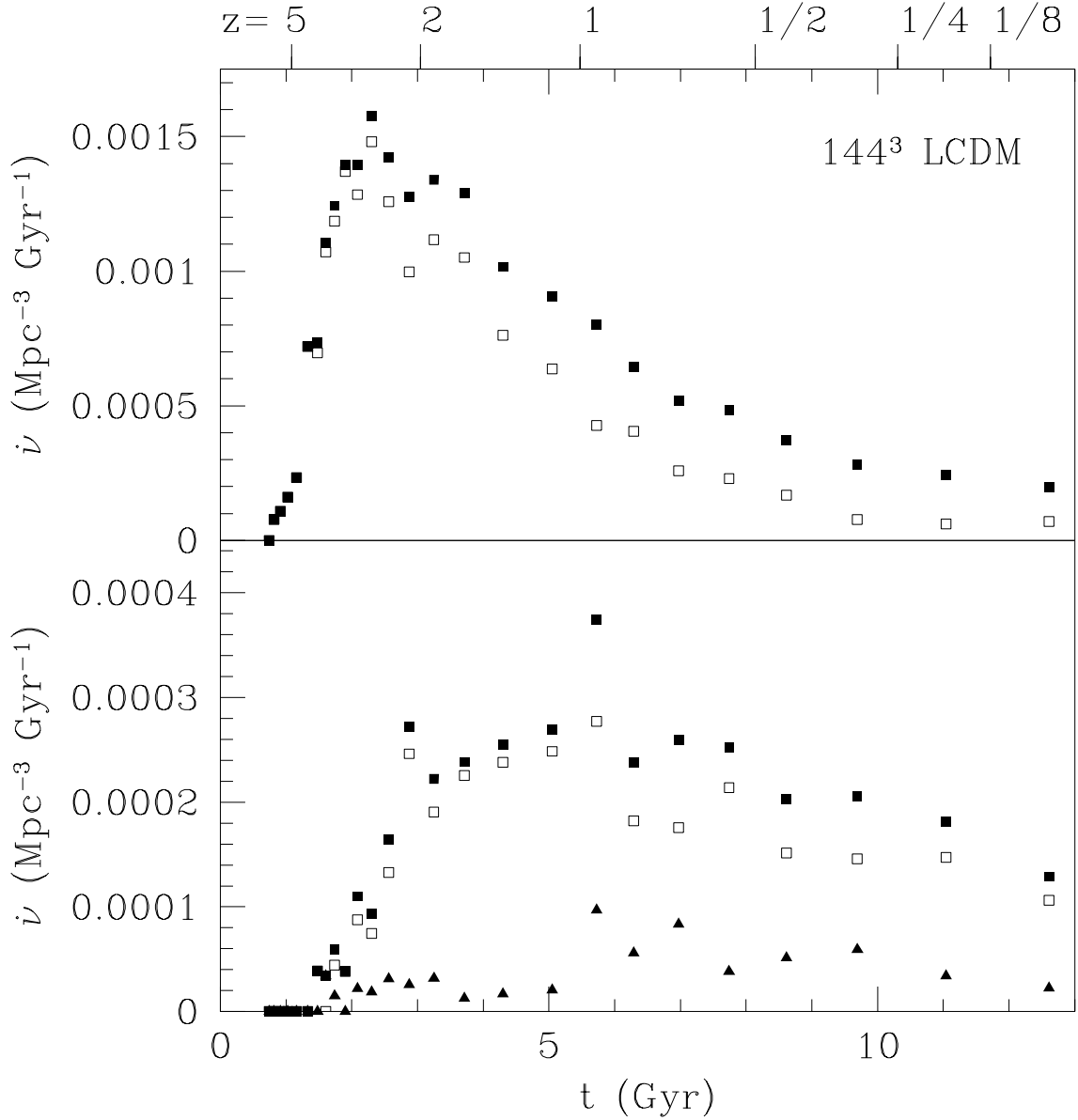}
\caption{
\label{fig.conrate144}
Contributions to the number density evolution
for galaxies with masses above $M_c = 5.4\times 10^{10} M_\odot$ (L50/144).
Top panel: filled squares give the creation rate of new
galaxies, and open squares give the net rate (creation minus destruction).
Bottom panel: filled squares give the total destruction rate, and
open squares and filled triangles show the respective contributions
of mergers and disruption.}
\end{figure*}
}
 
\newcommand{\acratefig}{
\begin{figure*}
\plotone{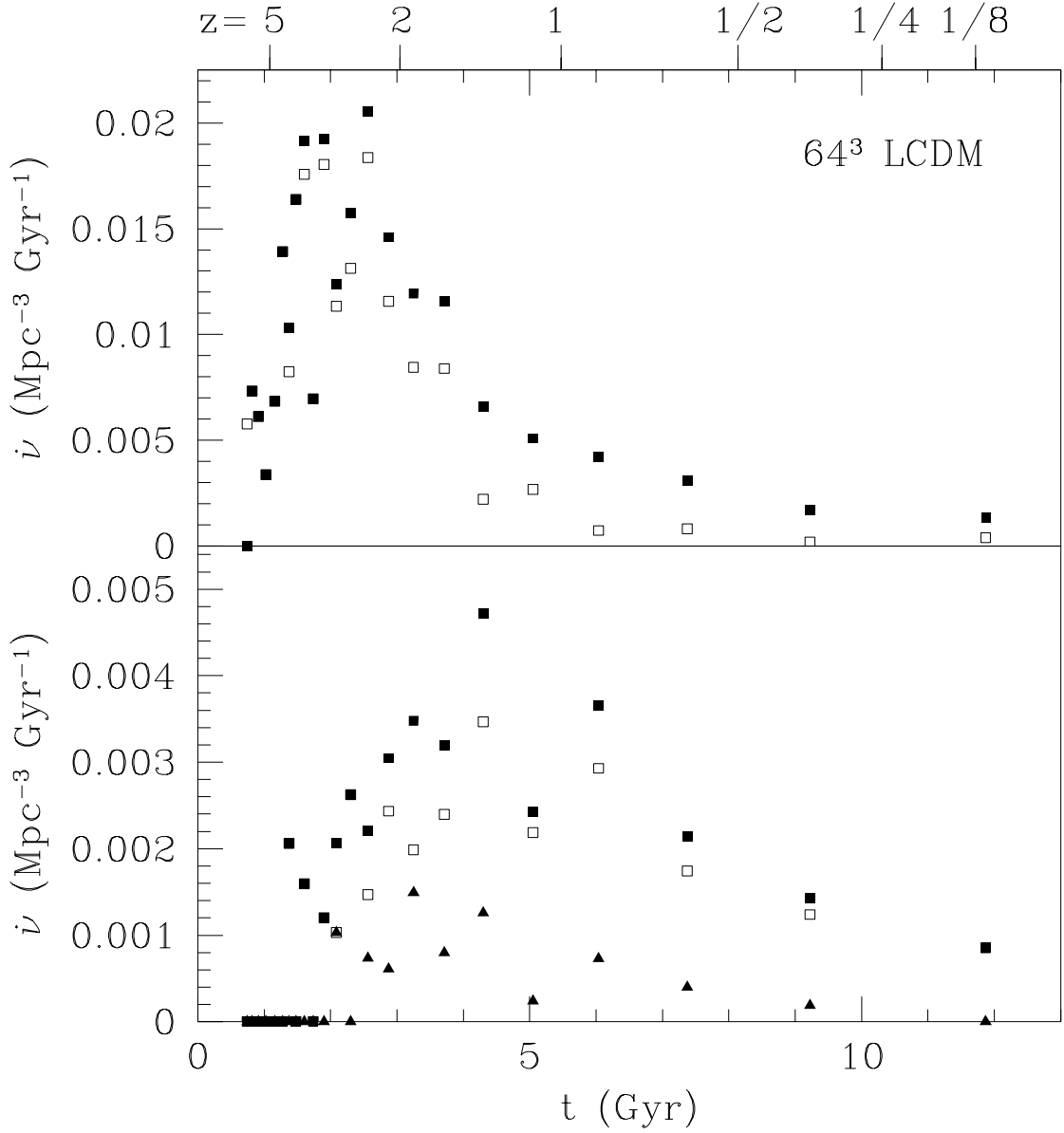}
\caption{
\label{fig.conrate64}
Same as Figure~\ref{fig.conrate144}, for a mass threshold
$M_c = 6.8\times 10^{9} M_\odot$ (L11/64).
Note the change in vertical scale, reflecting the
higher number density of the less massive galaxies.}
\end{figure*}
}

\newcommand{\massaccfig}{
\begin{figure*}
\plotone{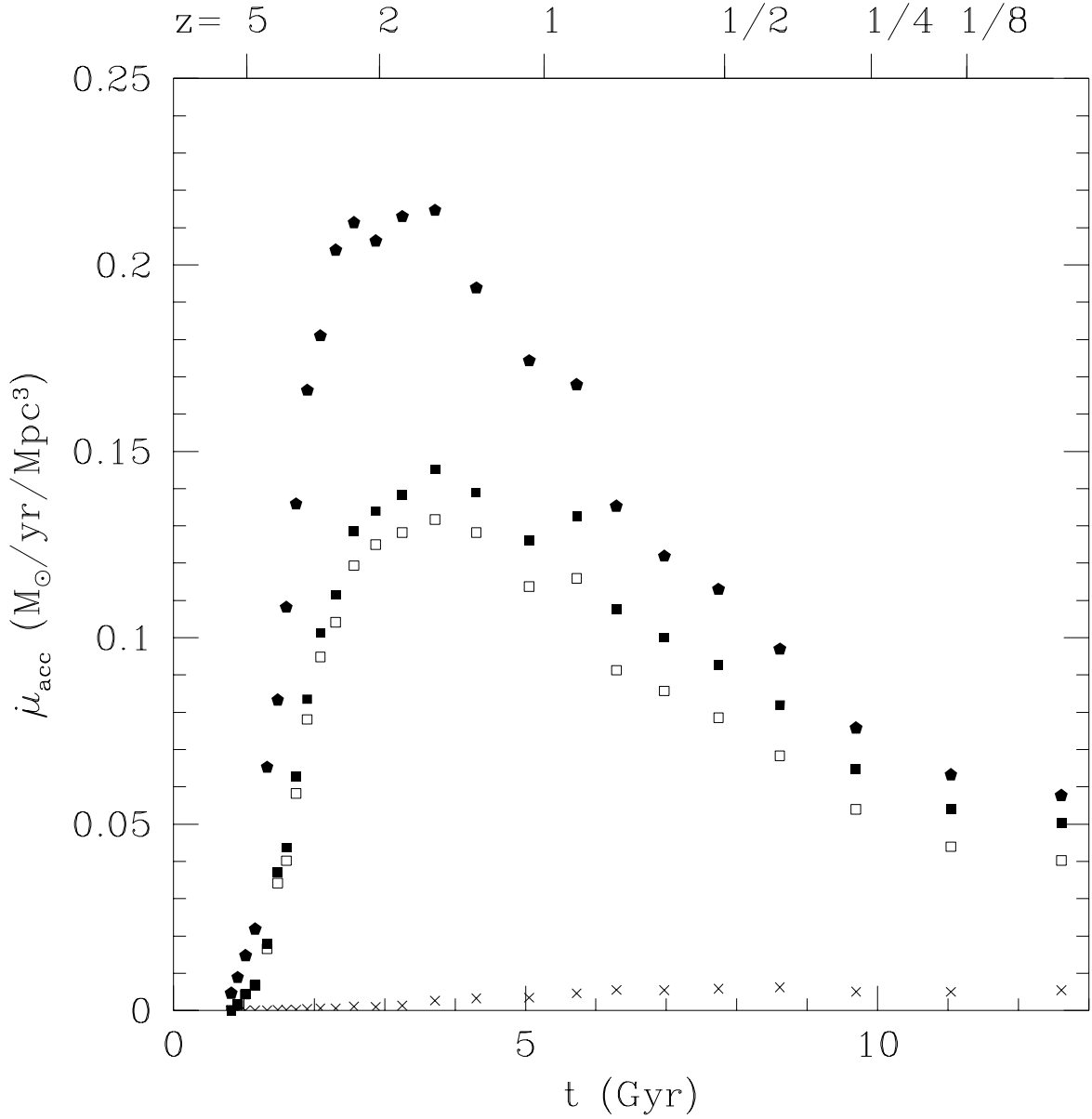}
\caption{
\label{fig.massacc}
Evolution of the 
mass accretion rate 
for galaxies above $M_c = 5.4\times 10^{10} M_\odot$ (L50/144).
solid pentagons show the total including the contribution
from newly formed galaxies.  Solid squares show only the
accretion rate onto existing galaxies, excluding newly formed objects.
Open squares show the contribution to this accretion from gas
and crosses the contribution from stars.}
\end{figure*}
}

\newcommand{\distaccfig}{
\begin{figure*}
\plotone{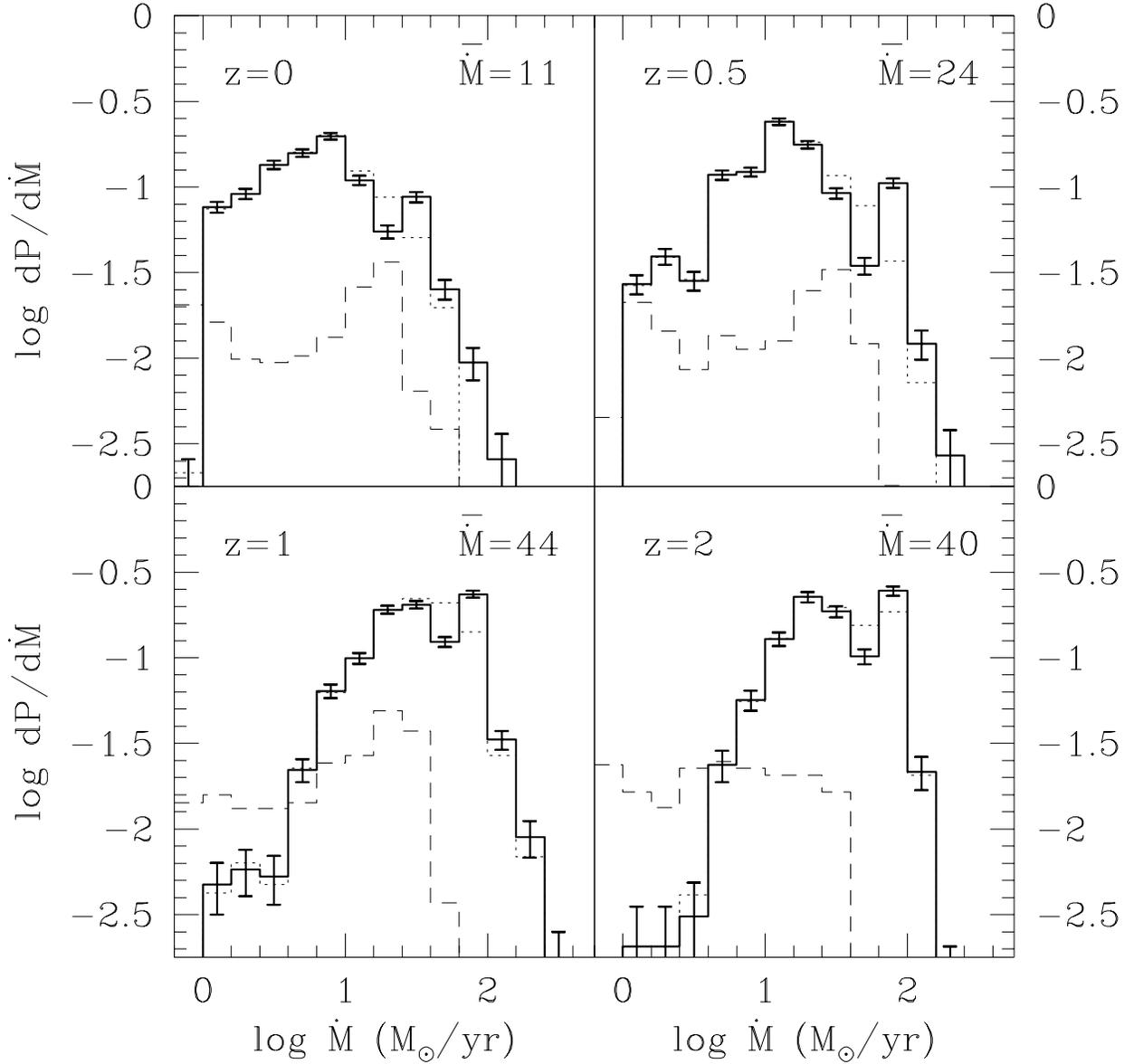}
\caption{
\label{fig.distaccfig}
Distribution of accretion rates onto individual galaxies at
different redshifts for galaxies with masses above 
$M_c = 5.4\times 10^{10} M_\odot$ (L50/144).  The solid line shows
total accretion rates, the dotted line gas accretion rates, and the
dashed line stellar accretion rates.  The quantity $\bar{\dot M}$
denotes the mean accretion rate at each time.}
\end{figure*}
}

\newcommand{\massmerfig}{
\begin{figure*}
\plotone{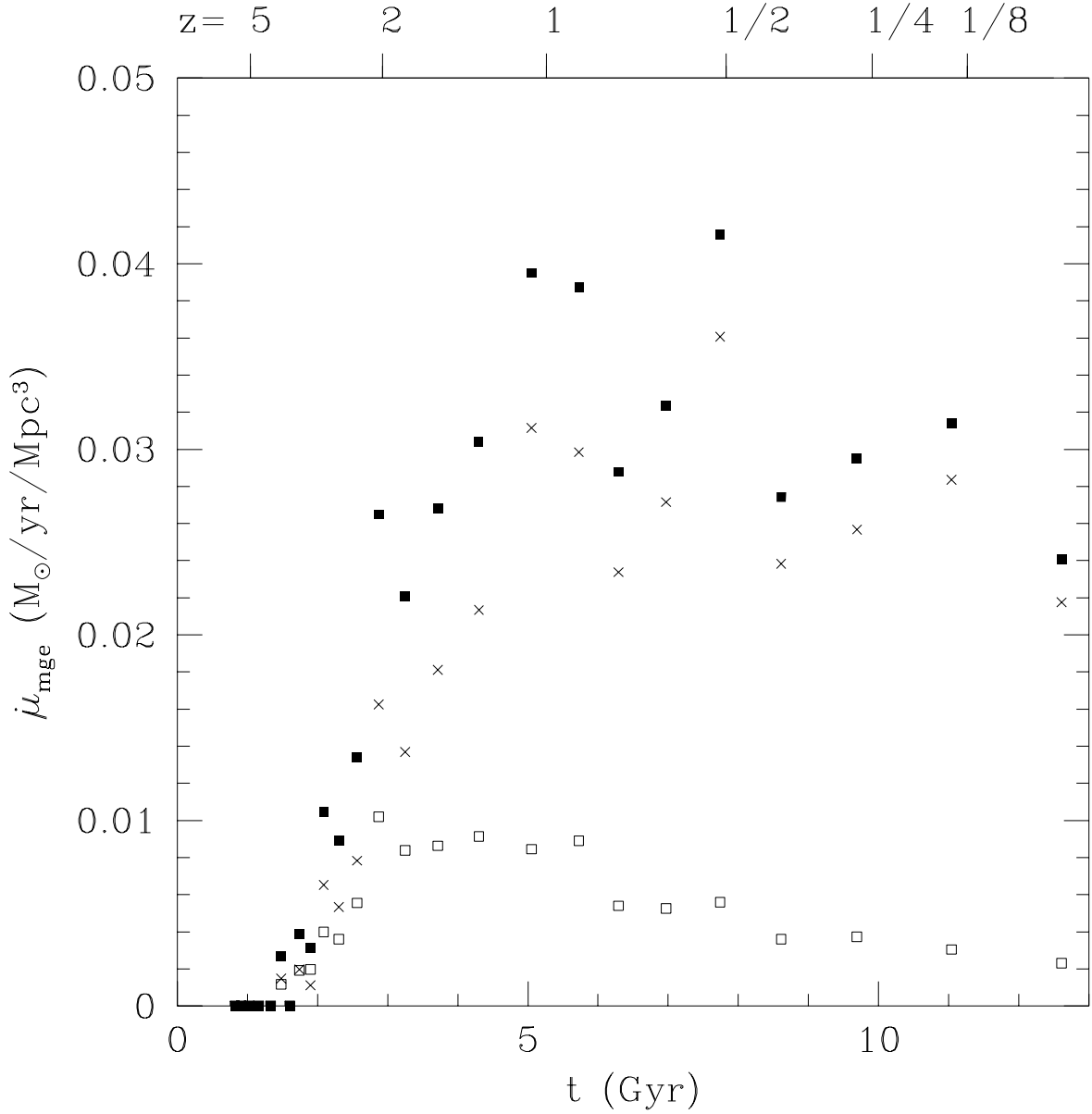}
\caption{
\label{fig.massmer}
The rate of mass gain through merging as a function
of time, for galaxies with masses above $M_c = 5.4\times 10^{10} M_\odot$ 
(L50/144).   Solid squares show the total rate, open squares the
contribution from gas, and crosses the contribution from stars. Note that
the vertical scale is stretched by a factor of five compared to 
Figure~\ref{fig.distaccfig}.}
\end{figure*}
}

\newcommand{\distmerfig}{
\begin{figure*}
\plotone{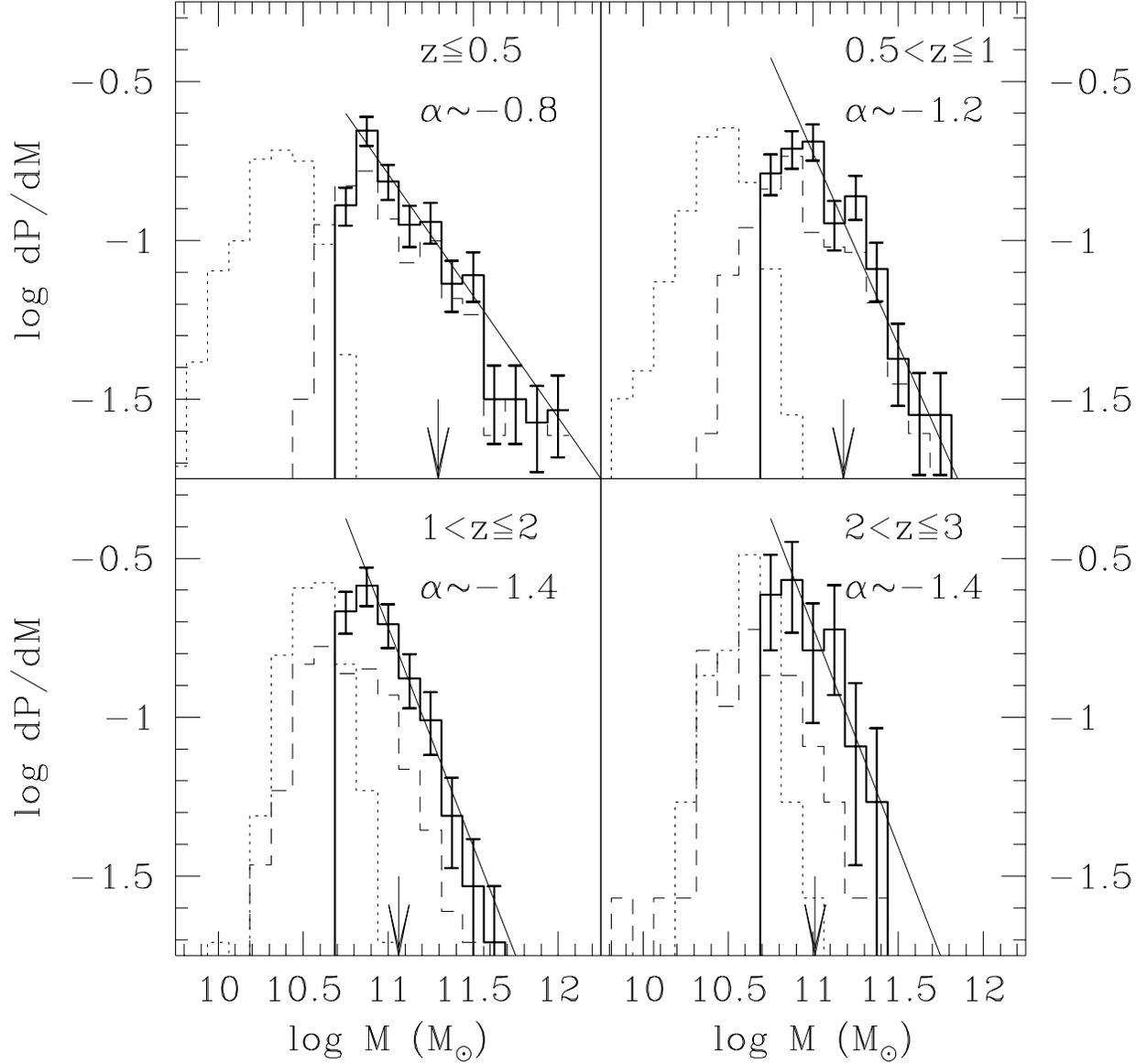}
\caption{
\label{fig.distmer}
The mass distribution of galaxies that merge into larger galaxies as a
function of smaller galaxy mass, in four redshift ranges, for galaxies
with masses above $M_c = 5.4\times 10^{10} M_\odot$ (L50/144).  The solid line
shows total merged mass, the dotted line the gas mass, and
the dashed line the stellar mass.  There is a cutoff in
total mass at $\log M=10.8$, corresponding to the minimum mass $M_c$,
although stellar and gas contributions can be
smaller.  The line indicates the approximate slope $\alpha$ of the
high-mass end.  The arrow on the abscissa denotes the mean merger
mass.}
\end{figure*}
}

\newcommand{\fracmassfig}{
\begin{figure*}
\plotone{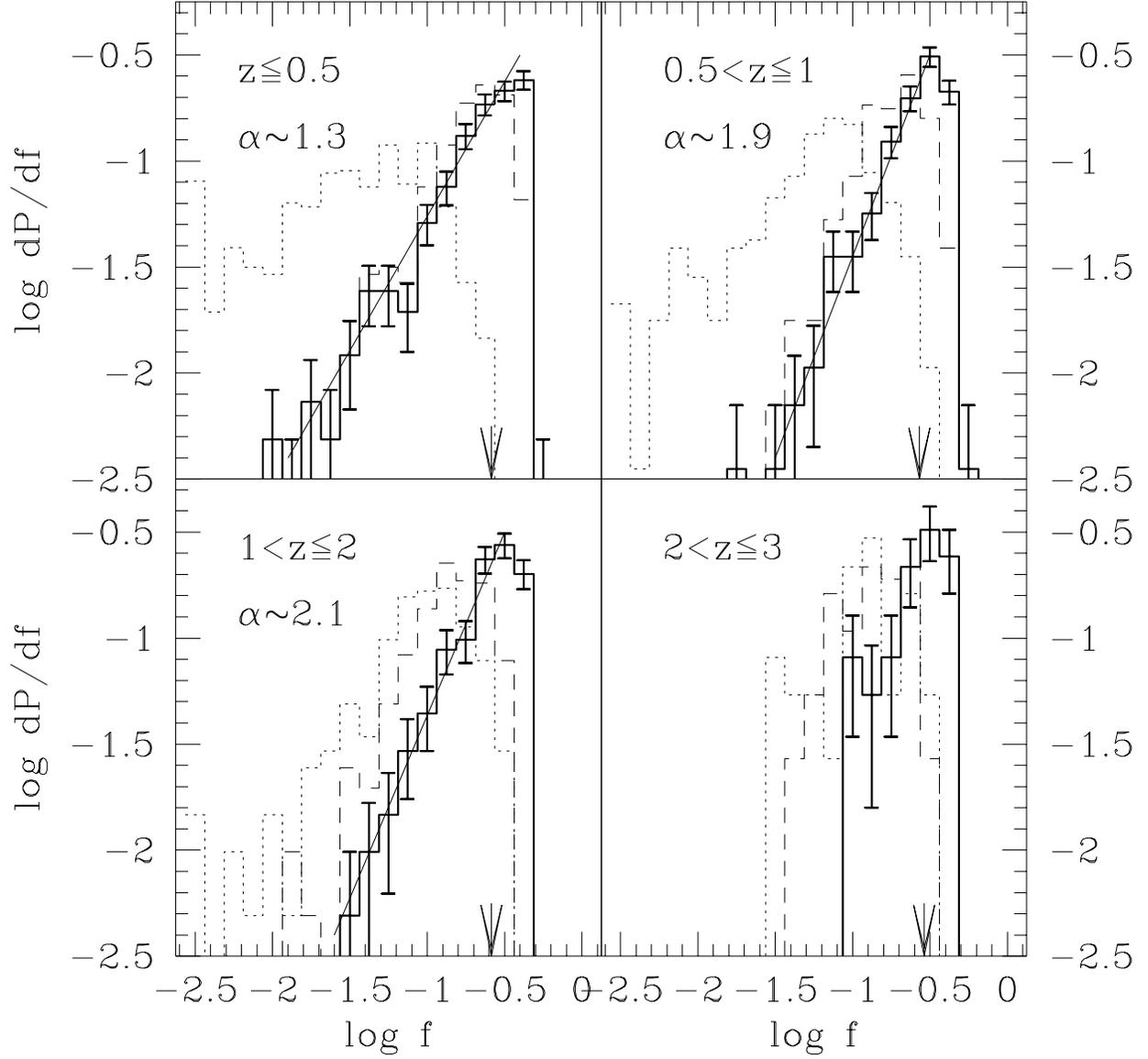}
\caption{
\label{fig.fracmass}
The fractional mass distribution of merger events into individual
galaxies as a function of merger mass ratio, $f = M_{sat}/M_{par}$,
in four different redshift ranges, for galaxies with masses
above $M_c = 5.4\times 10^{10} M_\odot$ (L50/144).  
The solid line shows the distribution of total fractional
mass, while the dotted and the dashed lines show the distribution of
the gaseous and the stellar mass fraction, respectively.  The line
indicates the approximate slope $\alpha$ of the distribution.  The
arrow on the abscissa indicates the mean mass ratio of the merger,
which is roughly $\bar{f}=0.25$ throughout.}
\end{figure*}
}

\newcommand{\numratefig}{
\begin{figure*}
\plotone{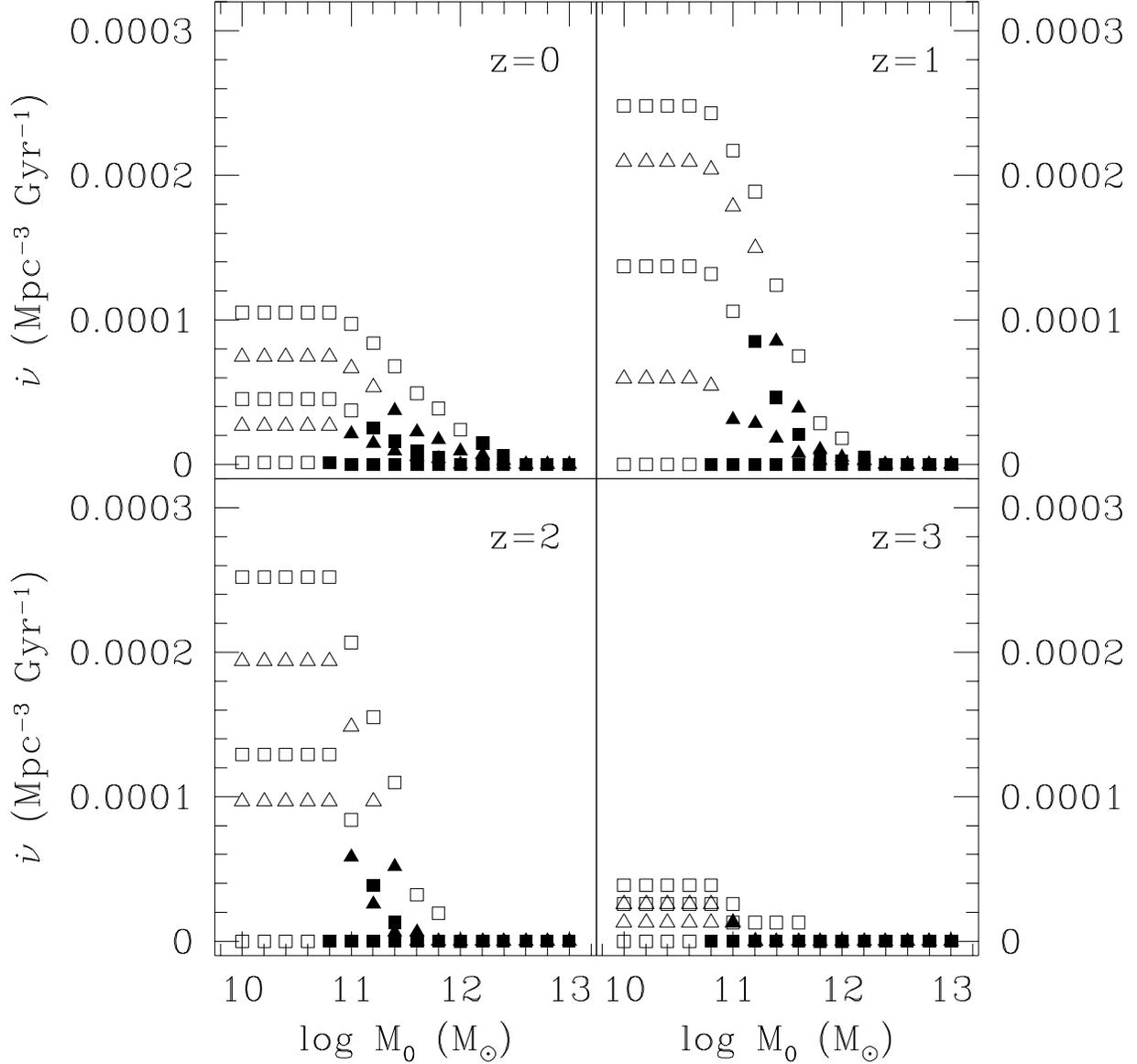}
\caption{
\label{fig.numrate}
The cumulative number merger rate as a function of parent group mass
for different merger ratios: $\dot{\nu}(\geq M_0,\geq f_0)$.  Each
sequence of points indicates a different $f_0$: from top to bottom,
$f_0=1\%,25\%,50\%,75\%,100\%$.  The solid symbols indicate the parent
mass range for which $M_{sat}\geq M_c$ for $f\geq f_0$, i.e. the mass
of the merged object lies above the 64-particle threshold.  The open
symbols indicate parent masses for which $M_{sat}<M_c$. Alternating
symbols are used for visual clarity.}
\end{figure*}
}

\newcommand{\massratefig}{
\begin{figure*}
\plotone{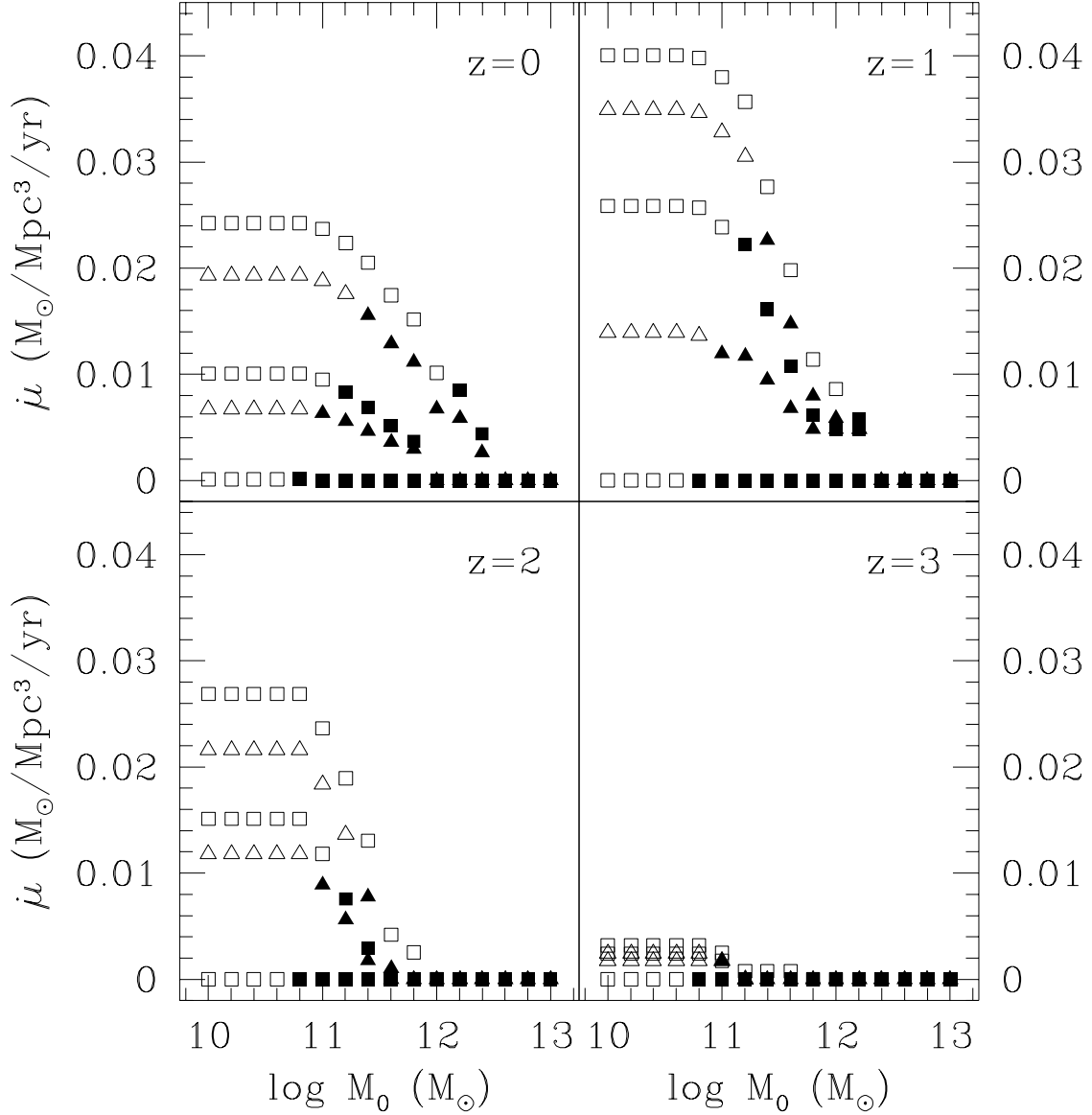}
\caption{
\label{fig.massrate}
The cumulative mass merger rate as a function of parent group mass for
different merger ratios: $\dot{\mu}(\geq M_0,\geq f_0)$.  Each
sequence of points indicates a different $f_0$: from top to bottom,
$f_0=1\%,25\%,50\%,75\%,100\%$.  The solid symbols indicate the parent
mass range for which $M_{sat}\geq M_c$.  The open symbols
indicate parent masses for which $M_{sat}<M_c$.  Alternating symbols
are used for visual clarity.}
\end{figure*}
}

\newcommand{\bigmergefig}{
\begin{figure*}
\plotone{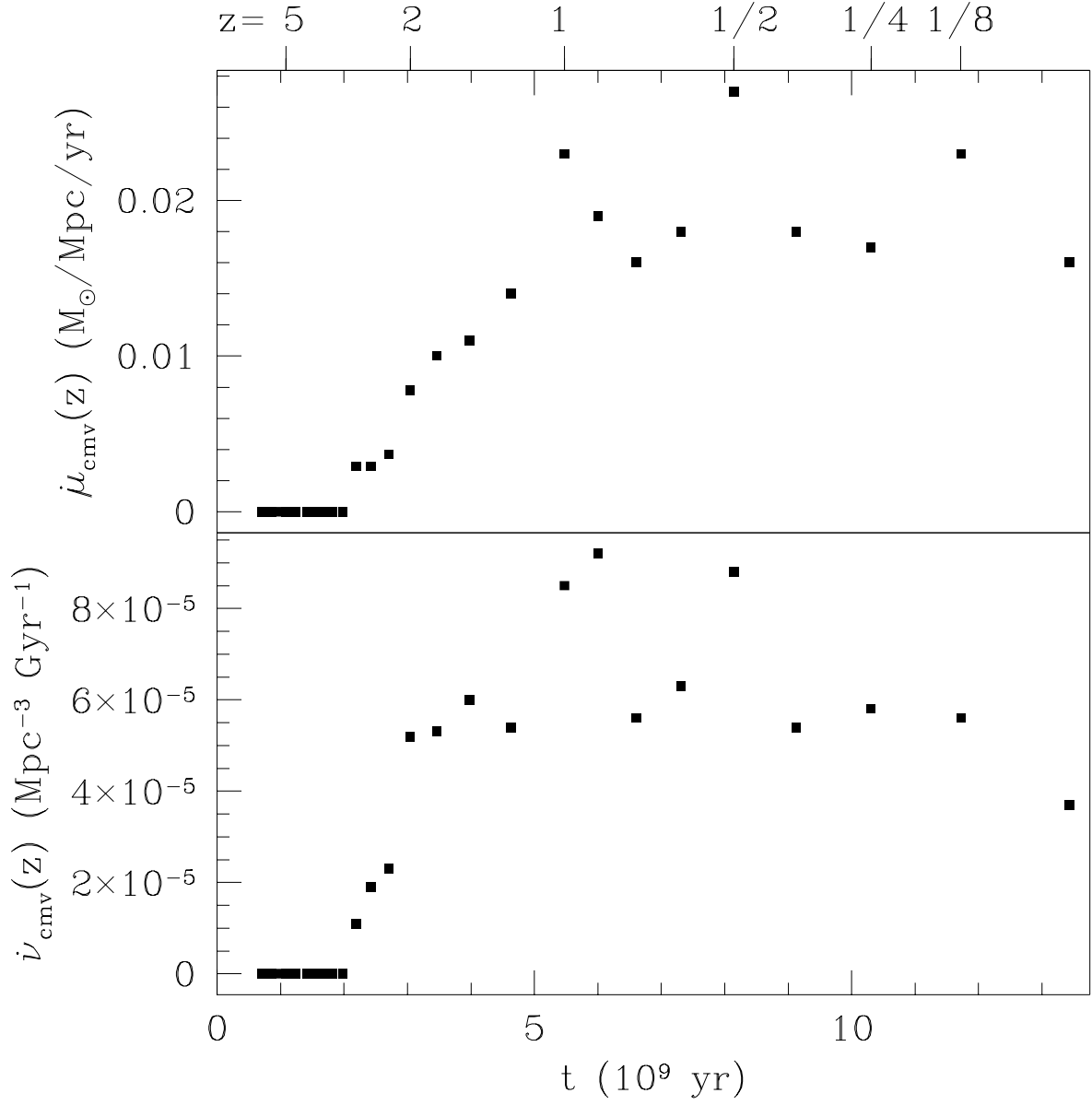}
\caption{
\label{fig.bigmerge}
The evolution of number (bottom) and mass (top) merger rates with
redshift for major mergers, $f_0=25\%$ and $M_0=2.2\times 10^{11}
M_{\odot}$.}
\end{figure*}
}

\newcommand{\lossfig}{
\begin{figure*}
\plotone{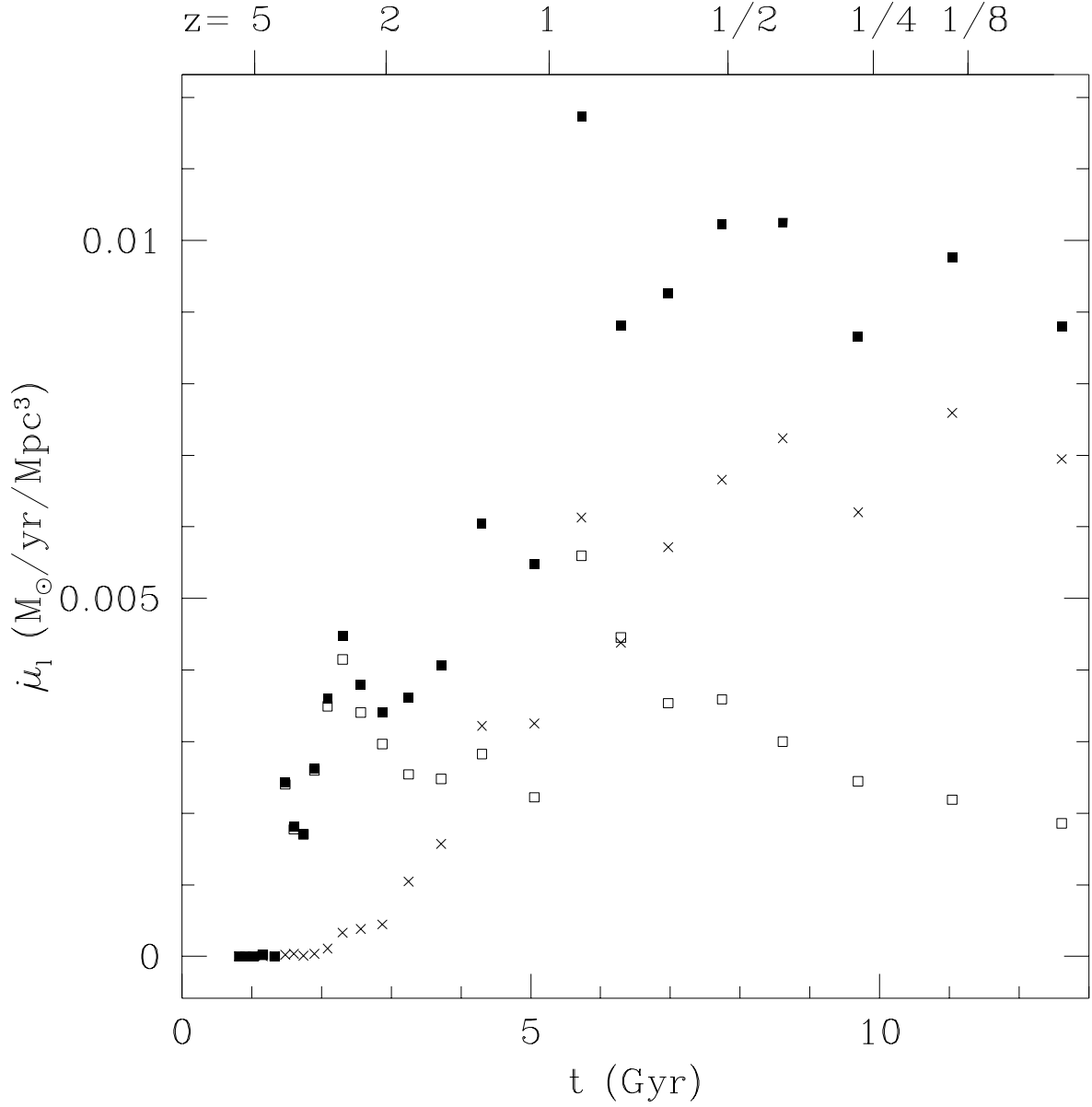}
\caption{
\label{fig.loss}
The rate of mass loss from existing groups per unit comoving volume as a
function of time, for galaxies above $M_c = 5.4\times 10^{10} M_\odot$ 
(L50/144).  Solid squares show the total, open squares
the contribution from gas, and crosses the contribution from
stars.}
\end{figure*}
}

\newcommand{\starfig}{
\begin{figure*}
\plotone{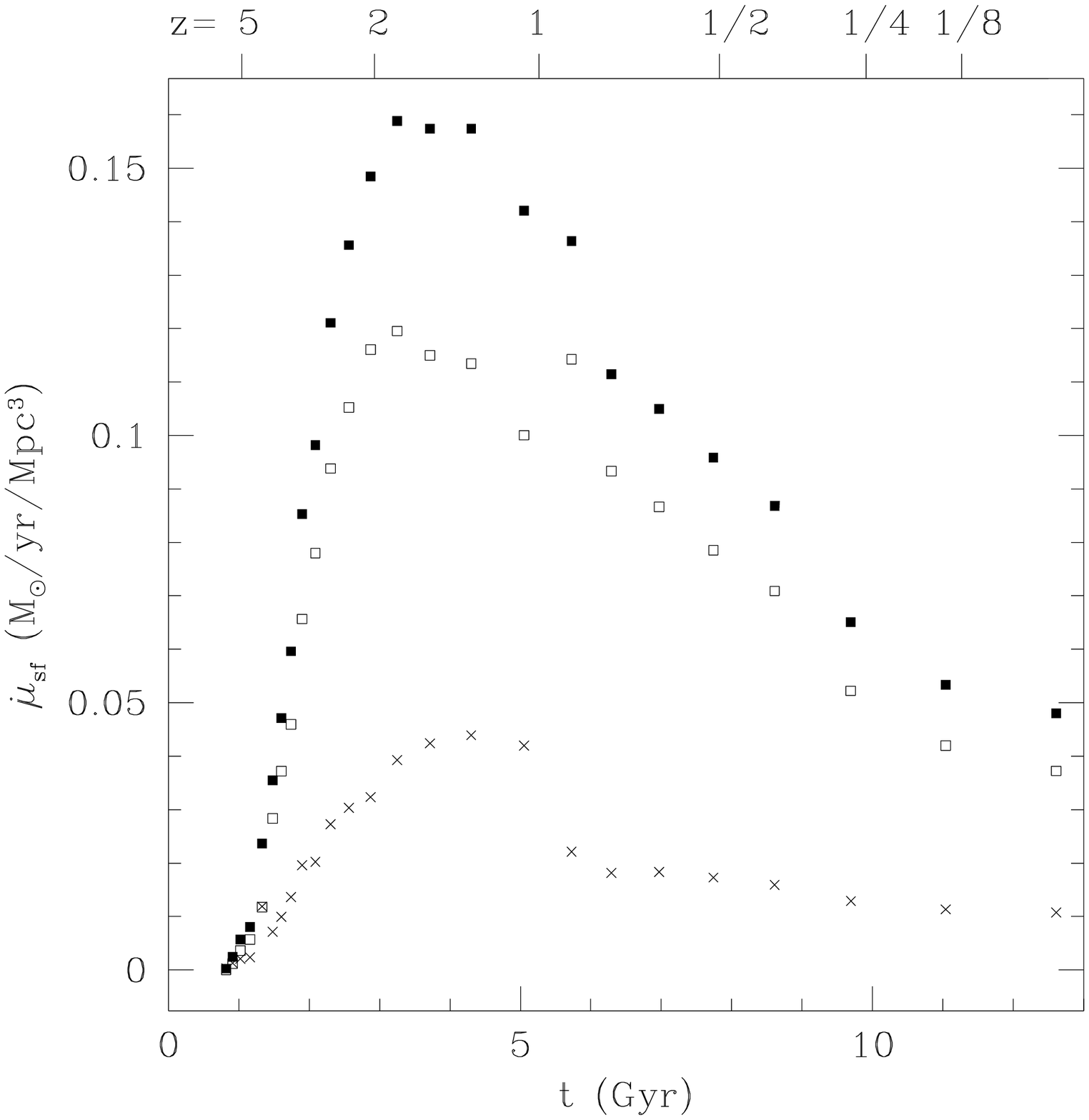}
\caption{
\label{fig.star}
The star formation rate in galaxies above $M_c = 5.4\times 10^{10} M_\odot$
(L50/144) per unit volume as a function of
time.  The solid squares show the total, the open squares the
contribution from gas already in groups and the crosses the
contribution from gas that is recently accreted.}
\end{figure*}
}

\section{Introduction}
What galaxies look like today clearly depends on how they were
assembled.  For example, the formation of ellipticals involves
considerably more random kinetic energy than does the assembly of
spirals.  Thus, it is thought that ellipticals frequently result from
the rather violent merger of roughly equal-sized galaxies (Toomre \&
Toomre 1972; Barnes \& Hernquist 1992).  To investigate this
possibility, observational studies have sought to find the merger
fraction as a function of redshift and compare this to the present
abundance of elliptical galaxies (e.g., Le F\`evre et al. 1999).
Conversely, the apparent quiescence of most field spirals implies
tight constraints on the amount of merging that can occur in
these systems and suggests that these galaxies more likely grew
through a process of smooth accretion (e.g., Toth \& Ostriker 1992).

Determining galactic histories and their relation to galaxy properties
is the essence of studying galaxy formation.  Theoretical investigations
of these histories must come to grips with a daunting array of
physical processes over a wide range of scales, all of which may
play important roles in molding a spectrum of initial density fluctuations
into a population of galaxies.
Hydrodynamic N-body simulations are a valuable tool because they
can represent many of these processes and scales, though they still
suffer from finite dynamic range and limited knowledge of the physics
of star formation.  In principle, one can use such simulations to
examine the link between the formation histories and morphological properties
of galaxies.  Unfortunately, simulations that resolve the internal
structure of galaxies and simultaneously model a representative 
volume of the universe are (slightly) beyond the reach of current computers. 
Furthermore, the most ambitious efforts to simulate the formation of
individual systems do not reproduce the observed distribution of
galaxy properties (e.g., Navarro \& Steinmetz 2000), and it is not clear
whether this failure has its roots in incorrect cosmological assumptions,
an inadequate treatment of physical processes (star formation and feedback
in particular), or numerical limitations of the simulations themselves.

In this paper we take a complementary approach to the study of galaxy assembly,
using simulations that represent large cosmological volumes but do not resolve 
the internal structure of galaxies.  Within these simulations we measure
the global evolution of the number density and total mass of the galaxy
population, and we examine the relative importance of smooth accretion 
and mergers in driving this evolution.  We also investigate some related
issues such as destruction of galaxies by tidal disruption, mass loss
from galaxies during mergers, and the connection between the global star
formation history and the history of galaxy assembly.
In the context of our adopted cosmological model, these numerical 
results provide a backdrop for interpreting observations of galaxy 
evolution and merger rates or constructing theoretical descriptions
of the origin of galaxy morphologies.
A combination of N-body simulations and analytic methods has led to
a fairly complete statistical understanding of the assembly of dark 
matter halos (see, e.g., \citealt{lacey93}; \citealt{sheth01});
our results represent an initial numerical effort to extend that understanding
to the baryonic components of galaxies.

This focused examination of the physics of galaxy assembly extends our 
earlier work on galaxy formation in the framework of the inflationary
cold dark matter scenario 
(Katz et al.\ 1992, 1996, 1999; Hernquist et al.\ 1995;
Weinberg et al.\ 1997, 1999, 2000). 
Here we take advantage of the larger simulations possible with the
parallel version of TreeSPH \citep{dave97}.
We also take advantage of an emerging consensus, driven by many
independent observations, that favors a cosmological
model with $\Omega_m\approx 0.4$, 
$h \equiv H_0 / (100 \kms \mpc^{-1}) \approx 0.65$,
a flat universe dominated by cold dark matter and vacuum energy,
and approximately scale-invariant primeval fluctuations with
the properties predicted by inflation.
This consensus allows us to concentrate our efforts on a single
set of cosmological parameters.  However, it is worth bearing in
mind that the viability of this model on galactic scales remains
a matter of debate (see, e.g., Moore et al.\ 1999; Spergel \& Steinhardt 2000),
and that even within its general framework there are uncertainties
in parameter values that could have a significant impact on its
predictions.

The principal limitation of our study is
that we cannot distinguish the process of truly smooth accretion
from mergers with galaxies below our simulations' mass resolution
threshold.  Throughout this paper, we will use the phrase ``smooth
accretion'' to refer to the combination of these two processes, as
distinct from mergers with objects above our resolution threshold.
For the large volume, high dynamic range simulation that we rely
on for most of our results, the space density of resolved objects
implies an identification with galaxies of luminosity
$\sim L_*/4$, where $L_*$ is the characteristic luminosity in
the \citet{schechter76} luminosity function (see discussion at
the end of \S\ref{sec:basic}.

The following Section describes our numerical simulations and our
method of identifying the dense groups of baryonic particles that
represent the observable regions of galaxies.  In \S 3 we investigate
the global evolution of the galaxy population --- number densities
and mass densities --- and we discuss numerical resolution issues
in some detail.  The heart of the paper is \S 4, where we examine
the processes that create and destroy galaxies in the simulations,
the roles of smooth accretion and mergers in galaxy assembly,
the statistics of merger masses and mass ratios, mass loss during
mergers, and the connection between star formation and accretion.
Section 5 discusses the implications of these results, the comparison
of our predicted merger rates to observations, and directions for
future work.  The Appendix presents
details of the analysis method, which is based on
a view of the simulations as Monte Carlo solutions of a kinetic
equation describing the evolution of the galaxy mass function.

\section{Numerical methods}
We perform our simulations using the parallel version of the
cosmological N-body/hydrodynamic code TreeSPH
(\citealt{hk89}; \citealt{kwh96}, hereafter KWH; \citealt{dave97}),
a code that unites smoothed particle
hydrodynamics (SPH) \citep{lucy77,gingold77} with the hierarchical
tree method for computing gravitational forces \citep{barnes86,h87}.
Dark matter, stars, and gas are all represented by particles;
collisionless material is influenced only by gravity, while gas is
subject to gravitational forces, pressure gradients, and shocks.  The
gas can also cool radiatively, assuming primordial abundances,
and through Compton cooling.  In SPH, gas properties are computed by
averaging or ``smoothing'' over a fixed number of neighboring
particles; the calculations here use 32-particle smoothing.  There is
a maximum allowed timestep, and all particles are integrated with this
step or one a power of two smaller.  The timestep criteria are
detailed further in KWH and \citet{quinn97}; we set
the tolerance parameter $\eta$ (defined in these papers)
to 0.4.  In these simulations we
include star formation and supernova feedback using the algorithm
described in KWH.  For each star formation event, supernova
energy is added to the surrounding particles as thermal energy on a
time scale of $2\times 10^7$ years.

\subsection{Simulation parameters}
\label{subsec:simulations}

All four of the simulations we discuss in this paper 
are based on a $\Lambda$-dominated 
cold dark matter (CDM) cosmological model with $\Omega_m = 0.4$,
$\Omega_\Lambda = 0.6$, 
$h \equiv H_0 / (100 \kms \mpc^{-1}) = 0.65$,
and primeval spectral index $n=0.93$.  With the tensor mode
contribution, normalizing to COBE using CMBFAST
\citep{seljak96,zaldarriaga98} implies a normalization $\sigma_8=0.8$,
which provides a good match to cluster abundances \citep{white93}.  We
use the \citet{hu96} formulation of the transfer function, and we adopt a
baryonic density $\Omega_b = 0.02 \, h^{-2}$, consistent with the
deuterium abundance in high redshift Lyman limit systems
\citep{burles97,burles98}, and the opacity of the high redshift
Lyman-alpha forest (e.g. Rauch et al. 1997).  
All of our simulations model a triply
periodic cubical volume.

The simulations cover a range of resolutions, which allows us to examine
the numerical behavior and test for convergence.  The primary and
largest simulation, L50/144 \citep{dave01}, has $144^3$ gas and dark
matter particles in a comoving periodic box $50\hinv\mpc$ on a side,
with a gravitational softening length $\epsilon_{grav} = 7 \, h^{-1}$
comoving kpc (equivalent Plummer softening). This simulation is
evolved to $z=0$.  As discussed in \S\ref{subsec:resolution} below,
we adopt as our nominal gas mass resolution the mass corresponding
to 64 SPH particles, which for this simulation is
$5.4 \times 10^{10} M_\odot$.  We also analyze
L11/64, a simulation with $64^3$ gas and dark matter particles in a
volume $11.11\hinv\mpc$ on a side, also evolved to $z=0$.  This
simulation has a particle mass eight times smaller than L50/144 and
a gravitational softening length two times smaller.
The minimum SPH smoothing length is $\epsilon_{grav}/4$, so the 
higher spatial resolution extends to the hydrodynamic forces.
We also analyze L11/128, a
simulation with $128^3$ particles of each type in a volume the same
size as the L11/64 simulation, but only evolved to $z=3$.  It has a
particle mass eight times smaller than L11/64 and a gravitational softening
length two times smaller.  Finally, we analyze L11/64$^\prime$. This simulation
has the same resolution and volume as L11/64, but it uses the same random 
realization of the CDM power spectrum as L11/128.  It has been evolved
to $z=3$, and the comparison to L11/128 gives our most direct test
of finite resolution effects down to this redshift.

\begin{table*}
\begin{center}
\caption{Simulations}
\begin{tabular}{ccccccc}
\hline
Name&$L (\mpc)$&$N$&$z_{fin}$&\multicolumn{1}{c}{$M_{max} (M_{\odot})$\tablenotemark{a}}&\multicolumn{1}{c}{$M_{min} (M_{\odot})$\tablenotemark{b}}&\multicolumn{1}{c}{$M_{c} (M_{\odot})$\tablenotemark{c}}\\
\tableline
L50/144&$50\hinv$&$2\times 144^3$&$0$&$3.1\times 10^{12}$&$6.8\times 10^9$&
                                        $5.4\times 10^{10}$\\
L11/64&$11\hinv$&$2\times 64^3$&$0$&$3.2\times 10^{10}$&$8.5\times 10^8$&
                                        $6.8\times 10^9$\\
L11/128&$11\hinv$&$2\times 128^3$&$3$&$3.0\times 10^{10}$&$1.1\times 10^8$&
                                        $8.5\times 10^8$\\
L11/64$^\prime$&$11\hinv$&$2\times 64^3$&$3$&$3.0\times 10^{10}$&$8.5\times 10^8$&
                                        $6.8\times 10^9$\\
\tableline
\end{tabular}
\tablenotetext{a}{Mass of 10$^{th}$ largest galaxy in simulation, at $z=0$ for 
L50/144 and $z=3$ for other simulations}
\tablenotetext{b}{Mass of 8-particle group}
\tablenotetext{c}{Mass of 64-particle group}
\end{center}
\end{table*}

Our results depend on the spatial and mass resolution of the simulation,
which we discuss in more detail below, and on the temporal spacing of the
available outputs.  Large time intervals tend to increase the amount
of smooth accretion because small groups can form and merge with
larger neighbors within a single time interval.  For our analysis, we
chose a sequence of outputs with a spacing of $\Delta z=0.5$ for $7\geq
z\geq 4$, $\Delta z=0.25$ for $4\geq z\geq 1$, and $\Delta z=0.125$ for
$1\geq z\geq 0$.  The corresponding time intervals, as seen in the
figures below, are roughly $1/3 \Gyr$, $2/3 \Gyr$ and $4/3 \Gyr$,
respectively.  These are on the order of the infall time scale at the
corresponding epoch, although perhaps somewhat longer at the latest
time.  We repeated some of our main analyses using twice as many
outputs, i.e. halving the redshift spacing, and found similar results.

\subsection{Group finding algorithm}
\label{subsec:skid}
SPH simulations with CDM initial conditions and radiative cooling
lead to the formation of dense groups of baryonic particles that
have sizes and masses comparable to the luminous regions of observed
galaxies (Katz et al.\ 1992; \citealt{evrard94}).
If star formation is included, these dense groups are the regions
where stars form (KWH).  The accretion and merger histories of these
objects are the subject of this paper, so the identification of distinct
particle groups underlies all of our subsequent analysis.
We use the group finding algorithm of Stadel et
al.\ (2001), Spline Kernel Interpolative DENMAX, abbreviated SKID (Gelb \&
Bertschinger 1994; see also KWH and
http://www-hpcc.astro.washington.edu/TSEGA/tools/skid.html). The basic algorithm
consists of: 1) determining the smoothed density field; 2) moving
particles upward along the gradient of the density field using an
heuristic equation of motion that forces them to collect at local
density maxima; 3) defining the approximate group to be the set of
particles that aggregate at a particular density peak; 4) finally,
removing particles from the group that do not satisfy a negative
energy binding criterion relative to the group's center of
mass.\footnote{This procedure starts with the highest energy particles
and updates the potential as particles are lost.}
In contrast to the identification of dark matter halos in N-body simulations,
there is essentially no ambiguity about the identification of 
distinct baryon clumps in these SPH simulations because dissipation 
greatly increases the density of cooled baryons with respect to the
local background.

As output, SKID produces a list of baryonic particles, both gas and stars,
that belong to each identified group.  To be included in a group, gas
particles must have a temperature $T<3\times 10^4\K$ and and a
density $\rho > 10^3\Omega_b$.  In effect, we consider only the cold
gas and stars, the material that comprises the bulk of the matter in
the central, visible regions of galaxies.  Therefore, throughout this
paper we use the terms {\it group} and {\it galaxy} interchangeably.

One complication that arises with this group definition is lost
particles.  Specifically, given all the particles in groups at one
output, SKID finds that roughly 1\% of these are lost by the next
output.  However, more careful inspection shows that roughly 30\% of
those particles declared lost still reside physically within the group
boundary, which we define to be the maximum radius of any bound member
particle.  These particles may in fact be unbound and in the process
of leaving the group, although we have not checked their energies
relative to the group center-of-mass.  Nevertheless, in the interest
of estimating the mass loss conservatively, we simply add these
particles back to the group.  Consequently, our group definitions
become dependent on output spacing, although the effect is only of
order 1\%.

\subsection{Mass resolution}
\label{subsec:resolution}

We allow SKID to identify groups with as few as eight particles.
Because these particles must be stars or cooled baryons, even these
small groups are real, high contrast objects residing in dense backgrounds.
However, our comparisons, statistical and direct, between simulations
with different mass resolutions show that a simulation's
group list becomes incomplete
(relative to a simulation with higher resolution) below $\sim 60$ particles.
Above this threshold, the total baryon mass of groups is well converged
(see the discussion of Figure~\ref{fig.massfunc} below),
though the division into stellar and gas components remains resolution
dependent up to a somewhat higher threshold.  
For most of the analyses of
this paper, we adopt $N_{grp}=64$ particles as the minimum number for a 
resolved group, and we discard all SKID groups with fewer particles.
Our nominal mass resolution limit is therefore that of 64 SPH particles,
though since mass is redistributed among particles during star formation
(see KWH) the exact mass of a 64-particle group may be slightly higher
or lower.  
We further discuss resolution effects and other limits on 
the mass ranges probed by our simulations in the next Section.

\section{Evolution of galaxy number and mass densities}
\label{sec:basic}
First we look at the simplest quantities that characterize the galaxy
populations in the simulations: the mass function, the total number
density, and the total mass density.  Examining these characteristics
shows qualitatively how the galaxy population evolves and also
illustrates the numerical factors that come into play in the analysis.

As mentioned above, the most important numerical variable is the mass
resolution, the minimum mass of a resolved particle group.
Because the number of galaxies increases towards low masses,
the number of galaxies in
a simulation depends sensitively on this lower mass limit.
The box size also comes into
play, however, as this determines the number of relatively rare,
high-mass objects that form, thereby determining the maximum
mass for which statistical results are meaningful.

Table 1 lists the numerical parameters of the simulations described
in \S\ref{subsec:simulations}, in particular the
characteristic minimum and maximum masses associated
with the resolution and box size. 
We define the maximum mass, $M_{max}$, to be the baryonic mass of the
10$^{th}$ largest galaxy in the simulation volume, identified
as described in \S\ref{subsec:skid}.  $M_{min}$ denotes the
mass corresponding to an 8-particle minimum group size, and $M_c$
denotes the adopted cutoff mass corresponding to the 64-particle
minimum.  
All three $11.11\hinv$ volumes have about the same
maximum mass at $z=3$.
The minimum mass is determined by the resolution, so the
L11/128 simulation has a substantially lower minimum mass.

Figure~\ref{fig.massfunc} shows particle groups in the L50/144, L11/64,
and L11/128 simulations at $z=3$.  In any given simulation, the mass
function begins to turn over at a mass 5-10 times higher than the
8-particle minimum of the SKID groups.  Comparison to the next higher 
resolution simulation shows that this turnover is a numerical artifact.
If we instead restrict attention to masses above our adopted
64-particle threshold $M_c$, then the mass function in
a higher resolution simulation gives a fairly smooth continuation of that in
the lower resolution simulation.  The envelope of the three mass functions 
approximately follows a Schechter function.

\ifthenelse{\boolean{apj}}{\massfuncfig}{}

All the analyses that follow include only galaxies with
masses above $M_c$.  
Because of their different resolutions and box sizes, our 
simulations probe different mass ranges.  In the case of
L50/144, one can see from Figure~\ref{fig.massfunc} that 
this range is nearly disjoint from that of the smaller volume simulations.

The lower panel of Figure~\ref{fig.density} compares the comoving number 
density $\nu$ of galaxies above the 64-particle threshold in the
four simulations, as a function of time (lower axis label) or redshift
(upper axis label).  The upper panel shows the comoving density $\mu$
of baryonic mass in these resolved galaxies.
Once again, the effects of mass resolution and box size are clearly
evident.  L11/128 has the highest number and mass densities.
L11/64 and L11/64$^\prime$ have fewer galaxies, of course, since
they have the same upper mass cutoff as L11/128 
but a lower mass cutoff that is eight times higher.
However, if the power law form of the low end of the mass function, roughly
$n(M)\propto M^{-1.1}$ in Figure~\ref{fig.massfunc}, continued to
arbitrarily high mass, then L50/144 would have the highest number
and mass densities because it has the largest dynamic range
$M_{max}/M_c$.  (For $n\propto M^{-1}$, the number of galaxies per
logarithmic interval is constant, and the mass per logarithmic interval
increases linearly with $M$.)  In reality, the maximum mass $M_{max}$
of L50/144 lies well into the exponential cutoff regime of the mass function,
and as a consequence the galaxy population of
L50/144 has the lowest mass density and by far the lowest number density of 
the four simulations.

\ifthenelse{\boolean{apj}}{\densityfig}{}

Turning from numerical considerations to evolution, we see that
number and mass densities both increase most strongly at early times,
i.e.\ for $z>2$.  In the simulations that continue to $z=0$, the number
of objects becomes approximately constant for $z<2$, and
the difference in mass density between L50/144 and L11/64 stays
about the same.  As discussed in \S\ref{sec:detail} below, the 
number of galaxies remains roughly constant even though merging
continues to late times because new galaxies also continue to form.
In contrast to the number density, the total mass in galaxies continues to
grow, both from the formation of new groups and from the accretion fed
growth of existing galaxies.

Figure~\ref{fig.numden} shows this behavior in terms of the
corresponding rates of change in mass density $\dot\mu$ 
and number density $\dot\nu$.
The differences between simulations are greatest
at high redshift, where the hierarchical mass
scale is lowest.  For $z<2$ the results at different resolutions agree,
indicating that the formation rate of very low mass objects has become
negligible.  The differences in total mass and number density apparent
in Figure~\ref{fig.density} must therefore be imprinted at early times.

\ifthenelse{\boolean{apj}}{\numdenfig}{}

The physical quantity one would like to know is the mass or number
density of galaxies above some specified mass $M$.
In practice $M$ cannot be below the mass resolution of
the simulation, $M_c$, and the mass and number densities measured by the
simulations
are not those for all galaxies above a given mass but only for galaxies below
a maximum mass determined by the volume.  However, we can measure the same
quantity in more than one of our simulations to assess the robustness of
our measurements given these limitations.  The L11/64 and L11/64$^\prime$ 
simulations have the same resolution
and box size, but there is a small difference in the quantities plotted
in Figures~\ref{fig.density} and \ref{fig.numden}, represented by the filled
squares and four point stars. These differences show the random fluctuations
associated with two different realizations of the initial power spectrum
in this volume.
We can assess the impact of our finite resolution on these
quantities by comparing the L11/64$^\prime$ simulation (four point stars)
to the L11/128 simulation analyzed with the mass cutoff 
$M_c=6.8 \times 10^9 M_\odot$ of L11/64$^\prime$.  These results are
shown by the
three point stars in Figures~\ref{fig.density} and \ref{fig.numden}.
The good agreement between the three and four point stars
indicates that we can robustly measure number and mass densities
for galaxies above the 64-particle threshold mass $M_c$, and that
finite resolution effects on our statistics above this threshold 
should be small.

To relate our numerical results to an observational context, we
would like to know the approximate luminosity of galaxies corresponding to
our resolution threshold $M_c$.  Although we have stellar masses
for each of our galaxies, the mass-to-light ratio of the stellar
populations depends sensitively on the assumed initial mass function
and on the age, metallicity, and extinction of the stellar population.
Furthermore, the overall scale of the
galaxy baryon masses may be sensitive to some cosmological parameters,
especially $\Omega_b$.  Given these uncertainties, the most robust
way to identify our mass threshold with an approximate luminosity
threshold is by matching the space densities of simulated and observed
galaxy populations.  If we match the mass function in 
Figure~\ref{fig.massfunc} to Blanton et al.'s (2001) determination
of the luminosity function from the Sloan Digital Sky Survey,
the implied luminosity corresponding to the threshold $M_c$
of L50/144 is $\sim L_*/4$, the precise value depending on whether
we match the $g^\prime$ or $r^\prime$ luminosity function and
whether we match space densities at $M_c$ or at $L_*$.

\section{Evolution of the galaxy population through accretion and mergers}
\label{sec:detail}

\subsection{Galaxy creation and destruction}
\label{subsec:creation}

The instantaneous rate of change in galaxy number in the simulations is
determined by the relative rates of formation and destruction.
Galaxies are destroyed either through merging with another larger
galaxy or though disruption.
In the upper panel of Figure~\ref{fig.conrate144}, open squares show
the rate of change $\dot\nu$ of galaxy number density from the L50/144 
simulation.  (These are the same as the open squares in the lower panel
of Figure~\ref{fig.numden}, but the vertical scale is greatly expanded.)
Filled squares show the creation rate of new galaxies; in general these
galaxies are not entirely new, but they have gained enough mass since
the last output to cross the mass threshold $M_c$.  The difference
between the filled and open points is the destruction rate, 
shown also by the filled squares in the lower panel.  Open squares
and filled triangles in this panel show the contributions to the destruction 
rate from mergers (which make one galaxy from two) and disruption (which
moves a galaxy from above $M_c$ to below $M_c$), respectively.
Figure~\ref{fig.conrate64} shows creation and destruction rates
for $M_c=6.8\times 10^9 M_\odot$, from L11/64.  This plot is noisier
because of the smaller number of galaxies in the simulation, but it
is qualitatively similar except for the higher number densities 
associated with less massive galaxies (note the change in 
vertical scale from Figure~\ref{fig.conrate144}).

In both cases the creation rate climbs rapidly to a peak at $z\approx 3$
and declines steadily thereafter, though new galaxies continue to form
(i.e., to cross the $M_c$ threshold) down to $z=0$.  The merger rate
climbs more gently, reaching a broad maximum at $z\sim 1-2$ and declining
only slowly at lower redshift.  Mergers always dominate over disruption
as a destruction mechanism.  At high redshift the creation rate is
much larger than the destruction rate, but the balance begins to shift
at $z<3$.  By the present day, the creation and destruction rates are
nearly equal, and both are much larger than the net rate.  
This result shows that a high galaxy merger rate need not lead to
rapid evolution of the galaxy luminosity function, since other galaxies
can grow to replace the ones that are lost.

\ifthenelse{\boolean{apj}}{\cratefig}{}

\ifthenelse{\boolean{apj}}{\acratefig}{}

\subsection{Growth of mass through accretion and mergers}
\label{subsec:growth}

Figures~\ref{fig.density} and~\ref{fig.numden} show that the total
mass in resolved galaxies increases with time, as one would expect.
Since we only count mergers among resolved objects, all new mass must
enter the population through accretion.  However, mergers redistribute
mass within the resolved galaxy population, and we can sensibly ask
whether existing resolved galaxies gain more of their mass through
accretion or through merging with other resolved systems.
Our methods for determining the smooth accretion and merger rates
from the simulation outputs are fully described in the Appendix.  
The most important point to recall here is that we define
accretion to be true smooth accretion plus any merging with galaxies
below the cutoff mass $M_c$, since we have no reliable way to distinguish
these processes.  For the remainder of
this paper we will restrict our discussion to the L50/144 simulation
because it has the largest dynamic range and probes the mass range of
greatest interest.

In Figure~\ref{fig.massacc}, pentagons show the total mass accretion
rate for galaxies with masses above $M_c = 5.4\times 10^{10} M_\odot$;
except for the minor effects of disruption and mass loss, this is the
same quantity shown by the open squares in Figure~\ref{fig.numden}
(top panel).  
Since we would like to know how {\it existing} galaxies acquire mass, we
subtract the contribution from newly formed galaxies and plot the
remainder as the filled squares.  
This globally averaged mass accretion rate in galaxies above
$5.4\times 10^{10}M_\odot$ rises rapidly until $z \approx 2$ 
and declines slowly thereafter.  At $z=0$, the accretion rate is
about $1/3$ of its peak value.  The rate of truly smooth accretion
by galaxies above $M_c$ is necessarily lower than that shown
in Figure~\ref{fig.massacc}, since higher resolution simulations
would resolve some of this accretion into mergers with small groups.
In addition to the total accretion rate, we show
the contributions from accreted gas (open squares) and stars (crosses)
separately.  As expected, gas dominates the type of accreted material.

\ifthenelse{\boolean{apj}}{\massaccfig}{}

Figure~\ref{fig.distaccfig} shows the distribution of accretion 
rates at four different redshifts.  The mean mass accretion rate
of resolved galaxies drops from $\dot M \approx 40 M_\odot{\rm yr}^{-1}$
at $z=1-2$ to $\dot M \approx 10 M_\odot{\rm yr}^{-1}$ at $z=0$,
though at each redshift the $\dot M$ distribution is broad.  Gas accretion
always dominates over stellar accretion, as expected from the global
properties in Figure~\ref{fig.massacc}.

\ifthenelse{\boolean{apj}}{\distaccfig}{}

Figure~\ref{fig.massmer} shows the volume averaged rate at which
resolved galaxies ($M> 5.4\times 10^{10} M_\odot$ in L50/144) gain
mass through merging with other resolved galaxies.
The single most important result of this paper comes from the
comparison of Figure~\ref{fig.massmer} to Figure~\ref{fig.massacc}:
galaxies gain most of their mass by accretion, not by mergers
(note the large change in vertical scale).
At $z \sim 2$, accretion dominates merging by about a factor of five.  
However, accretion declines more rapidly than merging towards 
low redshift, and by $z=0$ it dominates by only a factor of two.
The ratio of accretion growth to merger growth in galaxies above
this mass threshold would drop if our simulations had higher mass
resolution and could therefore resolve objects that are currently
counted in the smooth accretion rate.\footnote{However, a higher
resolution simulation would not necessarily give a lower 
accretion-to-merger ratio for its full population of resolved galaxies,
since its mass resolution threshold would also be lower, and these
less massive galaxies would still tend to accrete unresolved material.}
Nevertheless, the total merging rate is fully determined for 
galaxies with a mass above $M_c=5.4\times 10^{10} M_{\odot}$, and
in this mass range mergers contribute $\la 1/3$ of the
mass growth rate at every redshift.
Furthermore, we will show in \S\ref{subsec:mergermass}
that most of the merger contribution comes from relatively massive
objects, so the overall accretion-to-merger ratio would remain high
even with higher mass resolution (see discussion in \S\ref{sec:discussion}).

\ifthenelse{\boolean{apj}}{\massmerfig}{}

Focusing on the merger rate itself, we see that it climbs
fairly quickly until $z \sim 1.5$.  The variations thereafter 
appear largely stochastic, with a slight overall decline.
The clear decline in the number rate of merging (Figure~\ref{fig.conrate144})
is compensated by an increase in the typical mass of the merging
objects.  Open squares and crosses in Figure~\ref{fig.massmer} show the
separate contributions of gas and stars.  Gas-rich mergers are
important early on, where about half the accreted material is gas and
half is stars, but they are quickly overtaken by mergers with predominantly
stellar systems.

\subsection{Masses and mass ratios of merging galaxies}
\label{subsec:mergermass}

Figure \ref{fig.distmer} shows the probability distribution of the masses
of smaller ``satellite'' galaxies that merge into 
larger ``parent'' galaxies.  Above
the cutoff mass, $M_c$, the distribution generally appears to follow a
power law in mass.  The slope of this power law decreases with redshift,
presumably following the evolution of the non-linear mass scale.  At high
redshift ($z>0.5$), the estimated slope is steep enough ($\alpha<-1$)
that small objects dominate in number, while large objects dominate in mass
(i.e., an integral for the total mass of merging objects diverges
at the upper end).
However, at low redshift ($z<0.5$), the slope flattens, so that large
objects dominate in both number and mass.

\ifthenelse{\boolean{apj}}{\distmerfig}{}

Figure~\ref{fig.fracmass} shows the distribution of merger mass ratio,
the ratio $f=M_{sat}/M_{par}$ between the satellite galaxy
and the larger parent with which it merges.
At the earliest times ($z>2$), large mass ratios
predominate, although the total amount of merging is somewhat smaller
than at later times.  Subsequently, a low $f$ tail develops, and the
distribution becomes approximately scale free above the
effective resolution cutoff.  The probabilities
extend to lower and lower $f$ as the simulation evolves.  The slopes of
the distributions are quite steep, with higher mass
ratios ($f \ga 0.4$) dominating by number.  However, 
it is important to keep in mind that our finite resolution 
suppresses low-$f$ mergers, and since the number of
resolved galaxies is largest near $M_c$, there is some preference
for merger ratio near unity almost by definition.

\ifthenelse{\boolean{apj}}{\fracmassfig}{}

Given the resolution-limited mass, $M_c$, the simulation also gives an
exact or complete estimate of the total amount of merging above a
given mass ratio and corresponding parent mass.  Stated in more
mathematical terms, for a given merger ratio $f_0$, there is a
corresponding parent mass, $M_0=M_c/f_0$, for which the simulation
includes all mergers with $f\geq f_0$ and $M_{par}\geq M_0$.
Conversely, given $f_0$, the rate is not exact for $M_{par}<M_0$
because the total should include satellites with $M_{sat}<M_c$, which
are unresolved in the simulation.

Figure \ref{fig.numrate} shows the total number merger rate
$\dot{\nu}(\geq M_0,\geq f_0)$ at different redshifts, while Figure
\ref{fig.massrate} shows the corresponding mass merger rate
$\dot{\mu}(\geq M_0,\geq f_0)$.  Both figures display a range of
values of $f_0$ and $M_0$.  
Of particular interest is the amount of major merging, 
since that could transform
a galaxy's morphological type.  We define a major merger to be
$f\geq 0.25$.\footnote{Simulations show that for $f\sim 1$, mergers
essentially transform galaxy types (e.g. Barnes 1988, 1992; Hernquist
1992, 1993; Barnes \& Hernquist 1996), while for $f\sim 0.1$, galaxies
are damaged, but not completely destroyed (e.g. Fullager et al. 1993;
Walker et al. 1996).}  
With the present simulation, we can measure the
amount of major merging exactly for $M_{par}\geq M_0=2.2\times 10^{11}
M_{\odot}$, approximately $L_*$.  Overall, major merging 
appears to contribute significantly to the total amount of merging
measured in the simulation. 
For this threshold mass at $z\sim 1$, 50\% of the total
mass in merging comes from pairs with $f_0\geq 0.25$.  At $z=0$, the
contribution increases to nearly 75\%.  In terms of number, the rate
lies in the range from $10^{-6}-10^{-5} \Gyr^{-1}\mpc^{-3}$ for $z<3$.
Figure \ref{fig.bigmerge} shows the rates for both number and mass as a
function of redshift for major merging.
For $z>3$, major merging does not occur at
this resolution limit.  For $z<3$, both rates climb.  The number rate
reaches a stochastically varying plateau for $z<2$, while the mass rate
continues to climb until $z=1$, where it also reaches a
plateau.

\ifthenelse{\boolean{apj}}{\numratefig}{}

\ifthenelse{\boolean{apj}}{\massratefig}{}

\ifthenelse{\boolean{apj}}{\bigmergefig}{}

\subsection{Mass loss}
Fairly significant mass loss also occurs in the simulations.  Figure
\ref{fig.loss} shows the mass lost from galaxies as a function of
time for galaxies with masses above $M_c = 5.4\times 10^{10} M_\odot$ (L50/144).
The trend follows the merging history shown above --- it rises
until $z\sim 1$ and reaches a plateau for $z<1$.  Overall, mass loss
rates are about 25\% of merging rates.  Gaseous mass loss is nearly
constant with time, while stellar mass loss increases with time.
Stellar mass loss dominates after $z\sim 0.5$.  Visual inspection
shows that the mass loss occurs in highly clustered regions.

\ifthenelse{\boolean{apj}}{\lossfig}{}

\subsection{Star formation}
The volume-averaged star-formation rate is closely related to the growth of
galaxies in the simulation.  Figure \ref{fig.star} plots the global
star-formation rate $\dot{\mu}_{sf}$
for galaxies with masses above $M_c = 5.4\times 10^{10} M_\odot$ (L50/144)
as a function of time.
Interestingly, the overall rate (filled squares) closely follows the
global rate of mass accretion shown in Figure \ref{fig.massacc} rather
than the merging history shown in Figure \ref{fig.massmer}.  We also
plot separately the star formation contributed by gas that is newly
acquired through smooth accretion since the last time output and that
contributed by gas already
present within the parent galaxy or merged satellite in our last time
slice (crosses and open squares, respectively).  Most of the stars
form from gas already present in galaxies.  Moreover, since merging
rates are much smaller than star-formation rates (c.f. Figures 6 and
8), this implies that star formation typically occurs in gas that was
present in the parent galaxy at the previous timestep.  Hence, in our
simulations, gas that enters a galaxy typically waits for at least the 
interval between our outputs 
($1/3 \Gyr$ at high redshift and $4/3 \Gyr$ at low redshift) 
before converting to stars.

\ifthenelse{\boolean{apj}}{\starfig}{}

\section{Discussion}
\label{sec:discussion}
The results presented here describe the growth of the central
baryonic components of galaxies in the context of hierarchical
structure formation.  We have focused primarily on the relative
contributions of merging and smooth accretion to the rate at
which large galaxies gain mass.  Overall
we find that accretion dominates, especially at higher redshift,
while merging becomes progressively more significant at later times.

The principal limitation of our results is that
we cannot distinguish between smooth accretion
and merging with objects below our mass resolution threshold, which is
$M_c=5.4\times 10^{10} M_{\odot}$ for our primary simulation (L50/144).
As discussed at the end of \S\ref{sec:basic}, matching the observed
galaxy space density implies that this mass threshold corresponds
approximately to a luminosity $L_*/4$.  Our analysis gives well
defined, and, we believe, numerically robust predictions for the merger
rates of galaxies above this threshold with other galaxies above
this threshold.  These are lower limits to the merger rates of galaxies
above this threshold with all other galaxies, and our corresponding
accretion rates are upper limits to the rates of truly smooth accretion
by these galaxies.

The mass spectrum of merging galaxies shown in
Figure~\ref{fig.distmer} gives useful guidance to the possible
contribution of sub-resolution merging.
Over roughly a decade in mass, the mass spectrum is approximately
a power law $\propto M^\alpha$, with $\alpha \sim -0.8$ at
$z=0$ and $\alpha \sim -1.4$ at $z=1-3$.  
The cumulative distribution of mergers above mass $M$ scales
as $M^{\alpha +1}$, and the amount of mass contributed by
mergers above mass $M$ scales as $M^{\alpha +2}$, thus diverging
towards the high mass end of the spectrum.
Thus, the measured slopes of the merger mass spectrum indicate
that sub-resolution mergers should contribute relatively little
mass compared to resolved mergers.  To obtain a quantitative estimate,
we assume that the measured power law mass spectrum extends from
a high mass cutoff $M_h=6\times 10^{11} M_\odot$ (see Figure~\ref{fig.distmer})
down to a low mass cutoff $M_l=0$.  We then find that merging
galaxies below $M_c$ should contribute roughly 25\% of the total mass
in mergers for $z>1$ ($\alpha\sim-1.4$) and only about 5\% of the
total for $z<0.5$ ($\alpha\sim-0.8$).  Thus, at high redshift, the
simulation appears to account for more than 75\% of the total mass in
merging, whereas at low redshift it accounts for at least 95\%.  
Combined with the results in Figures~\ref{fig.massacc} and~\ref{fig.massmer},
this implies that merging overall accounts for no more than 25\% of
the total mass in accretion and merging combined for $z>1$ and no more
than 35\% of the total for $z<0.5$.
While higher resolution simulations will be needed to verify this
estimate, the extrapolation from our current results implies that 
truly smooth accretion always exceeds merging by at least a factor of
two in contributing to the mass evolution of galaxies.

It is interesting to compare the mass spectrum of mergers of baryonic
objects found here to the mass spectrum of dark matter halo mergers,
derived by \citet{lacey93} using the extended Press-Schechter formalism.
The comparison is not exact because we
determine the mass distribution of all merging objects, whereas 
\citet{lacey93} determine the distribution of objects falling into a
single halo.  The merger mass spectrum that they find is steeper than 
the one shown in our Figure~\ref{fig.distmer}.
Nevertheless, the high redshift behavior is qualitatively similar
in both cases: low mass objects dominate in number but high 
mass objects dominate in mass.
At low redshift there is a qualitative difference in the two results: our
scaling implies that high mass objects dominate in both
mass and number, while Lacey \& Cole's implies that low
mass objects still dominate in number but high mass objects dominate
in mass.  This difference could plausibly reflect the different 
dynamics of baryon and dark matter mergers.  When two dark halos
merge, the galaxies that they contain will not merge immediately.
Dynamical friction can drag together massive galaxies more quickly
than low mass galaxies, and this should make the merging mass spectrum of 
galaxies shallower than that of dark halos.

Our main point of contact with observations is through merging rates.
Observationally, the efforts focus on determining the
evolution of pair fraction with redshift (Zepf \& Koo 1989; Carlberg
et al. 1994; Yee \& Ellingson 1995; Woods, Fahlman, \& Richer 1995; Patton et
al. 1997; Abraham 1998; LeF\`evre et al. 1999; Carlberg et al. 2000).
The difficulty lies in identifying pairs of galaxies that should merge
in a time of order the system dynamical time at increasingly large
redshift.  Here we use the true physical merger rates, since these
are the quantities we have available from the simulations, but we do
not investigate whether the observational analyses accurately infer
these rates.

{}From their analysis of pairs selected from the CNOC2 and CFGRS
surveys, \citet{carl00} have recently estimated the rate of
mass growth due to merging in galaxies above about $0.2L_*$ out to
$z=1$.  They find an integrated mass growth rate of about $2\times
10^{-2}L_* \Gyr^{-1}$ per galaxy, with an estimated uncertainty
of a factor of two.  Their $0.2L_*$ threshold should
correspond approximately to our $M_c$ threshold, though there is
a factor $\sim 2$ uncertainty in this identification.
While \citet{carl00} do not
explicitly present mass ratios for their pairs, it is likely that for
such large galaxies the mergers are predominantly major ($f\geq
0.25$).  From Figure \ref{fig.bigmerge} we find that the major merging
rate $\dot\mu_{mge}$ at and above $M_c$ varies from $2\times 10^{-2}$
to $3.5\times 10^{-2} M_{\odot}\mpc^{-3}\yr^{-1}$ between $z=1$ and
$z=0$, while the comoving number density of objects varies from
$4.2\times 10^{-3}$ to $5.5\times 10^{-3}\mpc^{-3}$ over the same
interval.  Dividing the merger rate by the comoving number density
and by an assumed value of $2.2\times 10^{11}M_\odot L_*^{-1}$ yields
rates of mass growth from major merging in the range
$2\times 10^{-2}-5\times 10^{-2} L_* \Gyr^{-1}$ per galaxy.
The agreement with the \citet{carl00} results is encouragingly good,
given the theoretical and observational uncertainties that enter
the comparison.

Another interesting feature of our results is the mass loss due to
merging.  Although the group finding algorithm introduces
some uncertainty in the measurement, it does appear that a
significant fraction of the mass in merging satellites winds up as
intergalactic debris.  Visual inspection of the simulations
shows that this material
permeates galaxy clusters and large galaxy groups and forms low-density halos 
around
massive galaxies.  Recent observational and theoretical work has
provided evidence for the existence of such material, particularly in
galaxy clusters (e.g. Theuns \& Warren 1997; Ciardullo et al. 1998;
Ferguson, Tanvir \& von Hippel 1998; Calcan\'eo-Rold\'an et al. 2000).
In future work, we will examine in more detail the properties of the
debris produced by merging galaxies.

One obvious way to extend the work presented here is to employ
new, higher resolution simulations.  We are currently evolving
a simulation similar to L11/64 but using $128^3$ particles in
a $22.22\hinv\mpc$ box.  This simulation will have larger dynamic
range than L11/64 and greater overlap with L50/144, improving
our ability to assess numerical resolution effects.  In the longer
term, a simulation with $288^3$ particles in a $50\hinv\mpc$ box
would have the same mass resolution, $M_c=6.8\times 10^9M_\odot$,
as L11/64, sufficient to settle many of the remaining issues
regarding the growth of $L_*$ galaxies in numerical simulations.

Motivated in part by the kinetic equation approach described
in the Appendix, we have focused here on volume-averaged 
rates of accretion and mergers.  These rates provide a good global
view of activity in the resolved galaxy population, but there are
other questions that can only be addressed by examining accretion
and merger histories as a function of environment or galaxy by galaxy.  
For example,
we would like to know what fraction of galaxies experience a major
merger between $z=0.5$ and the present day, or what fraction have
quiescent accretion histories from $z=1$ to $z=0$, and we would like
to know how those fractions correlate with galaxy mass,
environment, and stellar age.  The analysis of mass acquisition
can also be extended to examine how galaxies gain angular momentum
or random kinetic energy.  We will present results from such
analyses in future work.  Studies of this sort,
applied to steadily improving numerical simulations, should provide
a solid foundation for understanding the mass, luminosity, and
morphological evolution of the galaxy population.

\acknowledgments

We thank Jeff Gardner for providing the basis for our merger code and
for performing the L11/128 simulation.  We also
thank Mark Fardal, Enrico Vesperini and Martin Weinberg for helpful
discussions.  This work was supported by NASA Astrophysical Theory
Grants NAG5-3922, NAG5-3820, and NAG5-3111, by NASA Long-Term Space
Astrophysics Grant NAG5-3525, and by the NSF under grants ASC93-18185,
ACI96-19019, and AST-9802568.  The simulations were performed at the
San Diego Supercomputer Center and NCSA.

\appendix

\section{Evolutionary equation}
The simulations analyzed above can be viewed as Monte Carlo solutions
of the kinetic equation that describes the evolution of the mass
function $n(M,t)$ (and includes a large number of additional degrees
of freedom, which we have projected over in our analysis).  In the
continuum limit, a simple form of the equation can be written:
\begin{eqnarray}
\pdrv{n}{t}&=&-\pdrv{}{M}\bigl[n \dot{M}\bigr]
	+\int_{M_c}^M dM' n(M-M',t)n(M',t)\Gamma(M-M',M',t)\nonumber\\
	&-&n(M,t)\int_{M_c}^{\infty} dM'n(M',t)\Gamma(M,M',t)
	+C(M,t)-D(M,t).
\end{eqnarray}
On the right-hand side, the first term gives the rate of change of the
baryonic mass $M$ of a galaxy with accretion rate $\dot M$.  Accretion
gives rise to advective evolution in the mass phase space so the term
is analogous to a continuity term.  In writing this, we have assumed
that the accretion $\dot M$ depends uniquely on the mass $M$, whereas,
in reality, a sample of galaxies at mass $M$ would have a distribution
of accretion rates.  What is written here can be thought of as the
evolution of the `average' galaxy of mass $M$.

The second and third terms describe the merging of galaxies.  The
former gives the rate of creation of new galaxies of mass $M$ through
the merging of pairs with masses $M'$ and $M-M'$ at the rate
$\Gamma(M-M',M,t)$.  The latter gives the loss rate of galaxies at $M$
through collisions with galaxies at $M'$ at the rate $\Gamma(M,M',t)$.
Note that we have ignored mass loss and disruption in writing these
terms.

The fourth and fifth terms denote the creation and destruction of
galaxies of mass $M$, respectively.  In the simulation, galaxies are
not created at all masses; instead they usually pass directly through
the minimum mass threshold $M_c$ by mass accretion.  In this case, one
could introduce the creation rate as a boundary condition.  Sometimes,
however, sub-$M_c$ galaxies can merge and produce a new galaxy with a mass
somewhat larger than $M_c$.  As a result, we can qualitatively
describe the creation function as a one-tail distribution that peaks
at $M_c$ and has a fairly short tail to higher mass.  Pure destruction
does not really occur in the simulations at our adopted mass threshold,
and the term has only been included for completeness.

By taking the first moment of equation (A1) with respect to $M$, one
obtains the equation describing the evolution of the total baryonic
mass in galaxies.  From this, one can see that the total mass can
change only through accretion onto and mass loss from existing
galaxies and through creation and destruction.  Merging, by contrast,
alters only the number of galaxies.

\section{Changes in total mass}
It is straightforward to generalize equation (A1) to the discrete form
that is required for analysis of the simulations.  Assuming the
discretized form of the equations, we write down the expressions
relating the total accretion and mass loss to the change in the total
amount of mass in galaxies between two times.  Let there be $N_1$
groups at time $t_1$ and $N_2$ groups at time $t_2$ where $t_2>t_1$.
Then the total mass in groups at either time $t_k$
\begin{equation}
M_k=\sum_i^{N_k} m_i,
\end{equation}
and the change in mass in groups between the two times
\begin{equation}
\Delta M=M_2-M_1=M_{acc}-M_{loss},
\label{eq:dM}
\end{equation}
where $M_{acc}$ is the mass accreted smoothly and $M_{loss}$ is the
mass lost from the progenitor groups at time $t_1$.  Here we have
included mass growth from created galaxies in the accretion term and
mass loss from destroyed galaxies in the loss term.  Defining $f_i$ as
the fraction of group $i$ at $t_1$ that contributes to some group at
$t_2$, we can write
\begin{equation}
M_{loss}=\sum_i^{N_1} (1-f_i) m_i,
\end{equation}
So that 
\begin{equation}
M_{acc}=M_2-(M_1-M_{loss})=M_2-\sum_i^{N_1}f_i m_i\equiv M_2-M_{grp},
\end{equation}
where $M_{grp}$ denotes the group mass at $t_1$ that contributes to
the group mass at $t_2$.  Using $M_{grp}$, we can define the merged
mass
\begin{equation}
M_{merge}=M_{grp}-\sum_j^{N_2} f_{i_{max}j} m_{i_{max} j},
\end{equation}
where $f_{i_{max} j}$ and $m_{i_{max} j}$ denote the fraction and
total mass of the largest group at $t_1$ that contributes to group
$j$ at $t_2$.  

\section{Changes in gas/star mass}
The expressions given above change when considering individual gas and
star components because of star formation.  In this case, defining
$M_i$ as the total gas mass in $N_i$ groups at time $t_i$, we can
write the total change in gas mass in groups:
\begin{equation}
\Delta M^g=M_2^g-M_1^g=M_{acc}^g+M_s^g-M_{loss}^g-M_g^s
\end{equation}
where $M_{acc}$ denotes the amount of gas accreted into groups,
$M_s^g$ the gas mass created from stars, $M_{loss}^g$ the amount of
gas lost from pre-existing groups, and $M_g^s$ the stellar mass
created from gas.  Of course, in the present simulations there is no
stellar evolution mass loss so $M_s^g=0$.  As in the definition above,
$M_{loss}^g=\sum_i(1-f_i) m_i^g$, the respective sum of the fraction of
mass in groups at $t_1$ that does not remain in groups at $t_2$ .
For stars, the treatment is analogous; however the star formation
terms change sign.  

To determine the $M_g^s$ term, we must consider the gas and star
masses of the individual particles at consecutive outputs.  However,
because of the finite time resolution, it is impossible to say whether
star formation occurs before or after material is added to the new
system.  Here we assume that star formation occurs after material is
added to a new system at time $t_2$.

With the total particle mass at time $t_1$
$\mu_{k,1}=\mu_{k,1}^g(t_1)+\mu_{k,1}^s$, the total stellar mass in
groups at $t_1$ that contribute to the total mass in groups at $t_2$ is
\begin{equation}
M_{grp}^s(t_1)=\sum_i^{N_1} f_i m_i^s=\sum_i^{N_1} f_i
	\sum_k^{n_i}\mu_{i,k,1}^s.
\end{equation}
However, the stellar mass of any particle may differ at $t_2$, so that
$M_{grp}^s(t_2)\neq M_{grp}^s(t_1)$.  The difference is, of course,
related to the net gain or loss due to star formation and stellar
evolution mass loss:
\begin{equation}
M_g^s-M_s^g=M_{grp}^s(t_2)-M_{grp}^s(t_1).
\end{equation}
Similarly, the change in gas mass implies that
\begin{equation}
M_s^g-M_g^s=M_{grp}^g(t_2)-M_{grp}^g(t_1).
\end{equation}
Since there is no stellar evolution mass loss, $M_s^g=0$.  The
treatment for accreted matter is analogous.

\ifthenelse{\boolean{apj}}{}{\clearpage}

\ifthenelse{\boolean{apj}}{}{\massfuncfig}

\ifthenelse{\boolean{apj}}{}{\densityfig}

\ifthenelse{\boolean{apj}}{}{\numdenfig}

\ifthenelse{\boolean{apj}}{}{\cratefig}

\ifthenelse{\boolean{apj}}{}{\acratefig}

\ifthenelse{\boolean{apj}}{}{\massaccfig}

\ifthenelse{\boolean{apj}}{}{\distaccfig}

\ifthenelse{\boolean{apj}}{}{\massmerfig}

\ifthenelse{\boolean{apj}}{}{\distmerfig}

\ifthenelse{\boolean{apj}}{}{\fracmassfig}

\ifthenelse{\boolean{apj}}{}{\numratefig}

\ifthenelse{\boolean{apj}}{}{\massratefig}

\ifthenelse{\boolean{apj}}{}{\bigmergefig}

\ifthenelse{\boolean{apj}}{}{\lossfig}

\ifthenelse{\boolean{apj}}{}{\starfig}

\end{document}